\documentclass[manuscript,nonacm]{acmart}

\usepackage{booktabs}
\usepackage{array}
\usepackage{makecell}
\usepackage{amsmath}
\usepackage{listings}
\usepackage{xcolor}
\usepackage{textcomp}
\usepackage{newunicodechar}

\newunicodechar{α}{\ensuremath{\alpha}}
\newunicodechar{θ}{\ensuremath{\theta}}
\newunicodechar{₁}{\textsubscript{1}}
\newunicodechar{₂}{\textsubscript{2}}
\newunicodechar{₃}{\textsubscript{3}}
\newunicodechar{−}{-}
\newunicodechar{∧}{\ensuremath{\wedge}}
\newunicodechar{✓}{[OK]}

\makeatletter
\def\maxwidth{\ifdim\Gin@nat@width>\linewidth\linewidth\else\Gin@nat@width\fi}
\def\maxheight{\ifdim\Gin@nat@height>.9\textheight.9\textheight\else\Gin@nat@height\fi}
\makeatother
\setkeys{Gin}{width=\maxwidth,height=\maxheight,keepaspectratio}

\definecolor{codegreen}{rgb}{0,0.6,0}
\definecolor{codegray}{rgb}{0.5,0.5,0.5}
\definecolor{codepurple}{rgb}{0.58,0,0.82}
\definecolor{backcolour}{rgb}{0.95,0.95,0.92}

\lstdefinestyle{mystyle}{
    backgroundcolor=\color{backcolour},
    commentstyle=\color{codegreen},
    keywordstyle=\color{magenta},
    numberstyle=\tiny\color{codegray},
    stringstyle=\color{codepurple},
    basicstyle=\ttfamily\tiny,
    breakatwhitespace=false,
    breaklines=true,
    captionpos=b,
    keepspaces=true,
    numbers=left,
    numbersep=5pt,
    showspaces=false,
    showstringspaces=false,
    showtabs=false,
    tabsize=2
}

\usepackage[ruled]{algorithm2e}

\SetAlFnt{\small}
\SetAlCapFnt{\small}
\SetAlCapNameFnt{\small}
\SetAlCapHSkip{0pt}

\providecommand{\tightlist}{%
  \setlength{\itemsep}{0pt}\setlength{\parskip}{0pt}}
\newcommand{\pandocbounded}[1]{#1}

\begin{document}

\title[Conformal Geometric Algebra as a Symbolic Interface for LLM-Driven 3D Scene Editing]{Conformal Geometric Algebra as a Symbolic Interface for LLM-Driven 3D Scene Editing}

\author{Manos Kamarianakis}
\email{kamarianakis@uoc.gr}
\affiliation{%
  \institution{University of Crete}
  \city{Heraklion}
  \country{Greece}
}
\affiliation{%
  \institution{FORTH -- Institute of Computer Science}
  \city{Heraklion}
  \country{Greece}
}
\affiliation{%
  \institution{ORamaVR SA}
  \city{Lausanne}
  \country{Switzerland}
}

\author{Pandelis Sofianos}
\affiliation{%
  \institution{ORamaVR SA}
  \city{Lausanne}
  \country{Switzerland}
}

\author{George Papagiannakis}
\affiliation{%
  \institution{University of Crete}
  \city{Heraklion}
  \country{Greece}
}
\affiliation{%
  \institution{FORTH -- Institute of Computer Science}
  \city{Heraklion}
  \country{Greece}
}
\affiliation{%
  \institution{ORamaVR SA}
  \city{Lausanne}
  \country{Switzerland}
}

\renewcommand{\shortauthors}{Kamarianakis, Sofianos, and Papagiannakis}
\renewcommand{\authorsaddresses}{%
Manos Kamarianakis (Corresponding author), kamarianakis@uoc.gr, University of Crete, Greece and FORTH -- Institute of Computer Science, Greece and \mbox{ORamaVR SA}, Switzerland; Pandelis Sofianos, \mbox{ORamaVR SA}, Switzerland; George Papagiannakis, University of Crete, Greece and FORTH -- Institute of Computer Science, Greece and \mbox{ORamaVR SA}, Switzerland.%
}

\begin{abstract}
What symbolic format should an LLM emit to perform reliable 3D scene editing from natural language, and does algebraic structure contribute beyond compact syntax? We study this question using Conformal Geometric Algebra (CGA) as a compact symbolic interface, contrasted with a verbose Euclidean 4$\times$4 matrix baseline and a non-CGA Compact SE3 control, in a natural-language 3D scene editing pipeline coupling controlled prompting with deterministic geometric execution.

Our primary finding concerns compositional fidelity under sequential instruction chains: in a sequence-stress protocol (20 templates, 6 trials each; \texttt{n=120} outputs per method), Simple CGA and Compact SE3 both reach 100\% parse validity, but Simple CGA preserves exact ordered operation chains more reliably (97.5\% vs 90.0\%, two-proportion \texttt{p=0.016}) at lower completion-token cost (112.6 vs 133.6 tokens), suggesting lower generation burden for compositional editing. This pattern is consistent with the hypothesis that algebraic expression form supports compositional faithfulness beyond compactness alone.

A secondary finding is confirmatory in the powered hard semantic suite (\texttt{n=100} per method): compact representations (Simple CGA: 45.0\%, Compact SE3: 42.0\%, Shenlong: 44.0\%) all exceed the Euclidean 4$\times$4 baseline (24.0\%), with Simple CGA vs Euclidean at +21 pp (\texttt{p=0.0028}) and Compact SE3 vs Euclidean at +18 pp (\texttt{p=0.0103}). The Simple CGA vs Compact SE3 contrast is statistically close (\texttt{p=0.7755}). A methodological contribution is the explicit separation of parse validity from geometric correctness, which reveals substantial optimization headroom invisible to syntax-only metrics.

Taken together, the results suggest that compact symbolic interfaces are a primary driver of reliability-cost gains, and that CGA's motor composition formalism may provide an additional advantage for ordered instruction chains. These findings have direct implications for real-time natural-language editing in immersive and interactive 3D environments.
\end{abstract}


\maketitle

\section{Introduction}\label{1-introduction}

The intersection of Large Language Models (LLMs) and 3D spatial
reasoning presents a fundamental challenge: how can a language model,
trained on sequential token prediction, reliably generate mathematically
precise geometric transformations? In natural-language-driven immersive
environments, this challenge is compounded by interaction latency:
sub-100ms round-trip targets constrain interface design, and every token
the model must generate translates directly into wall-clock delay in the
editing loop. Choosing the right symbolic output representation is
therefore not only a correctness decision but a real-time viability
decision for interactive systems.

Recent work has explored using LLMs for robotics control, scene graph
manipulation, and code generation for 3D environments, yet a persistent
bottleneck is the \emph{representation} through which the LLM
communicates geometric intent to the execution engine.

The dominant approach encodes transformations as 4×4 homogeneous
matrices, requiring the LLM to generate 16 numerical entries with
correct trigonometric values for rotations, proper displacement vectors
for translations, and consistent diagonal entries for scaling. This
representation is verbose, error-prone, and poorly matched to the
token-generation strengths of language models. A rotation by 90$^\circ$ around
the Z-axis, for instance, requires 32 tokens in matrix form versus 6
tokens as a CGA rotor expression \texttt{R(pi/2,\ e1,\ e2)}.

Conformal Geometric Algebra (CGA) offers a principled alternative. By
embedding 3D Euclidean space \(\mathbb{R}^3\) into the conformal model
\(\mathbb{R}^{4,1}\), rigid motions (translations and rotations) and
uniform scalings (dilations) are represented as \emph{motors} applied
through the unified sandwich product
\(P' = M \cdot P \cdot \widetilde{M}\). This algebraic unification
eliminates the need for separate code paths for different transformation
types and, crucially for LLM integration, produces dramatically more
compact symbolic expressions.

Kamarianakis and Papagiannakis~\cite{ref1} introduced the Shenlong system,
demonstrating that CGA-based prompting yields 16\% faster LLM responses
and 9.6\% higher success rates compared to their matrix-based baseline
on spatial reasoning tasks. In this paper, we extend that line of work
with an explicit comparative framing: compact symbolic interfaces are
evaluated against a Euclidean 4×4 baseline, with CGA analyzed as one
high-expressiveness compact representation rather than as a uniquely
privileged endpoint. Our contributions are:

\begin{enumerate}
\def\labelenumi{\arabic{enumi}.}
\item
  \textbf{Systematic comparison across compact-vs-verbose
  representations.} We benchmark three prompting strategies --- the
  original Shenlong verbose CGA prompt, a streamlined Simple CGA prompt,
  and a Euclidean 4×4 baseline --- across 33 experiments spanning
  simple translations, compositional multi-operation tasks, scaling
  experiments from 5 to 100 objects, and spatial accuracy tests; and we
  add a Compact SE3 control in hard semantic follow-up.
\item
  \textbf{Representation-compactness analysis.} We provide empirical
  evidence that compact symbolic outputs are associated with stronger
  reliability-cost profiles than verbose matrix outputs: in the powered
  full-loop latency protocol (Section~\ref{57-interactive-latency-and-engage-pilot}), Simple CGA shows lower median
  latency than Euclidean (1.19s vs 1.40s, \textasciitilde210ms gap),
  while Compact SE3 is lower still (0.81s median); in sequence-stress,
  compact methods sustain lower output burden than Euclidean while
  preserving high parse validity.
\item
  \textbf{Fully reproducible implementation.} A reference implementation
  is provided (Section~\ref{appendix-b-implementation-details}), using the Kingdon library~\cite{roelfs2025willingkingdoncliffordalgebra} for CGA
  computations, with deterministic template engines for known spatial
  patterns, GPT-4o-mini integration for novel instructions, and
  multi-view 3D visualization for spatial validation.
\end{enumerate}

Methodologically, the paper evaluates four questions: reliability at
scale, stability under repeated runs, semantic validity beyond parse
success, and the role of representation versus prompt/policy factors.
Claims are intentionally protocol-conditional for the tested setup and
are not presented as universal performance guarantees. Where compact
non-CGA and compact CGA methods are statistically close, we report that
proximity directly and avoid uniqueness claims.

The remainder of this paper is organized as follows.
Section~\ref{2-related-work} reviews related work in LLM-based 3D
manipulation, geometric algebra, and structured output approaches.
Section~\ref{3-mathematical-foundations-conformal-geometric-algebra}
presents the mathematical foundations of CGA in the Cl(4,1) algebra.
Section~\ref{4-system-architecture} describes the system architecture,
including the scene representation, the three prompting strategies, and
the CGA execution engine.
Section~\ref{5-experiments-and-results} presents experimental results
ordered by evidential strength, leading with the confirmed
sequence-fidelity finding.
Section~\ref{6-discussion} discusses the findings, analyzes improvement
opportunities, and outlines limitations with explicit threats to
validity.
Section~\ref{7-conclusion} concludes with future work directions.

\begin{center}\rule{0.5\linewidth}{0.5pt}\end{center}

\section{Related Work}\label{2-related-work}

\subsection{LLMs for 3D Scene Understanding and
Manipulation}\label{21-llms-for-3d-scene-understanding-and-manipulation}

The application of LLMs to spatial reasoning has evolved rapidly. Early
work by Liang et al.~\cite{ref2} demonstrated that LLMs could generate robot
control code from natural language, while subsequent systems like
SayPlan~\cite{ref3} and VoxPoser~\cite{ref4} explored LLM-driven planning for
embodied agents. These approaches typically output either high-level
plans (requiring a separate controller) or imperative code snippets
(requiring a full simulation API).

A key working hypothesis in CGA-oriented approaches is that a
\emph{declarative algebraic representation} --- specifying \emph{what}
transformation to apply rather than \emph{how} to implement it --- is
better suited to LLM generation. The LLM need only produce a compact
expression like \texttt{T(3*e1)*R(pi/2,e1,e2)}, and the algebra engine
handles the rest.

\subsection{Geometric Algebra in Computer
Graphics}\label{22-geometric-algebra-in-computer-graphics}

Geometric Algebra (GA) has a long history in computer graphics and
robotics~\cite{ref5,ref6}. Dorst et al.~\cite{ref7} provided foundational
treatments of GA for geometric computing, while Hildenbrand~\cite{ref8}
developed efficient implementations. Conformal Geometric Algebra has
been applied to inverse kinematics~\cite{ref9} and rigid body dynamics
~\cite{ref11}. In parallel, educational and framework-oriented graphics work
has supported broader GA adoption~\cite{ref10,ref12}.

The MAGES platform~\cite{ref12} has integrated GA-based transformations for
medical XR training, demonstrating the practical applicability of
algebraic methods in production systems. The Shenlong system~\cite{ref1}
is, to our knowledge, an early direct integration of CGA with LLM-based
reasoning for 3D scene editing, establishing the paradigm that this paper
extends.

\subsection{Structured Symbolic Interfaces for Spatial
Reasoning}\label{23-structured-symbolic-interfaces-for-spatial-reasoning}

Hybrid neural-symbolic pipelines combine neural language understanding
with symbolic structure~\cite{ref14}. In this paper, "symbolic" refers to
executable geometric structure rather than logic-level symbolic
inference. CGA is one candidate symbolic layer because it is both
mathematically rigorous (providing algebraic guarantees on
transformation correctness) and syntactically compact (enabling reliable
LLM generation). This motivates CGA as a \emph{candidate interface}
where the LLM provides language understanding while the algebra provides
executable geometric structure.

This interface-level framing also motivates the constrained-decoding
literature reviewed next.

\subsection{Structured Output and Constrained
Decoding}\label{24-structured-output-and-constrained-decoding}

A parallel line of work addresses LLM output reliability through
syntactic enforcement rather than representation design.
Grammar-constrained decoding frameworks --- including LMQL~\cite{ref15},
Outlines~\cite{ref16}, and Guidance~\cite{ref17} --- restrict the
model\textquotesingle s token sampling to strings that satisfy a
user-specified formal grammar, guaranteeing parse-valid outputs by
construction. Provider-level structured output APIs (e.g.
OpenAI\textquotesingle s JSON Schema mode) apply similar constraints at
inference time, ensuring that returned JSON conforms to a declared
schema without requiring post-hoc validation.

These approaches address a real bottleneck: in our own ablation results
(see Section~\ref{58-ablation-diagnostics}, Table~\ref{tab:17}), the Shenlong CGA variant reaches only 80--82\%
parse success under standard sampling, a failure mode attributable to
prompt-length-induced truncation rather than representational
difficulty. Grammar constraints would eliminate such failures entirely
for any representation family.

However, syntactic enforcement and representation choice operate at
different levels of the pipeline and are not substitutes. A constrained
decoder can guarantee that the model emits a well-formed 4×4 matrix or a
valid CGA expression string, but it cannot enforce that the emitted
values are \emph{geometrically correct} --- that trigonometric entries
are consistent, that displacement magnitudes match the intended spatial
relationship, or that an operation sequence preserves ordered
compositional intent. The parse-semantic gap documented in Section~\ref{55-semantic-validity-under-harder-grounding},
where all methods maintain near-perfect parse rates while semantic
correctness plateaus at 25--45\%, arises precisely in this
post-syntactic regime that format constraints do not reach.

CGA\textquotesingle s relevance in this context is therefore distinct
from a formatting argument. The motor formalism encodes translation,
rotation, and dilation under a single algebraic operator family, meaning
that a correctly expressed CGA motor is geometrically correct by
construction --- the algebra engine guarantees transformation
consistency without numerical checking. A correctly expressed 4×4
matrix, by contrast, still requires that the 16 emitted entries
collectively satisfy rotation group constraints and displacement
coherence, conditions that grammar-constrained decoding does not
enforce. The representation choice thus determines how much geometric
correctness the model must supply versus how much the execution engine
can guarantee, a distinction that grammar-constrained decoding leaves
unresolved.

Closely related is recent work on learned scene languages. SceneScript
~\cite{ref18} learns a structured language for 3D scene representation from
data, enabling LLM-compatible scene serialisation without hand-designed
algebraic primitives. SpatialLM~\cite{ref19} and StructDiffusion~\cite{ref13}
similarly ground language models in structured spatial representations.
These approaches demonstrate the value of compact,
language-model-compatible spatial encodings, but rely on learned
representations that require large training corpora and offer less
algebraic interpretability than analytic motor formulations. Our work is
complementary: we study the regime where analytic geometric guarantees
are desirable and the symbolic interface must be designed rather than
learned, as is the case in applications requiring verifiable
transformation correctness or integration with existing geometric
algebra runtimes such as the MAGES platform~\cite{ref12}. In our setting,
correctness is verified directly by the deterministic execution engine
rather than by a learned oracle.

\begin{center}\rule{0.5\linewidth}{0.5pt}\end{center}

\section{Mathematical Foundations: Conformal Geometric
Algebra}\label{3-mathematical-foundations-conformal-geometric-algebra}

\subsection{The Conformal Model
Cl(4,1)}\label{31-the-conformal-model-cl41}

Conformal Geometric Algebra extends 3D Euclidean space \(\mathbb{R}^3\)
with two extra basis vectors to create the conformal model
\(\mathbb{R}^{4,1}\). The algebra Cl(4,1) has signature (4,1), meaning
four basis vectors that square to +1 and one that squares to −1,
yielding a \(2^5 = 32\)-dimensional algebra with basis blades from grade
0 to grade 5.

The five basis vectors are:

\begin{itemize}
\tightlist
\item
  \textbf{Three Euclidean basis vectors:} \(e_1\) (X, right), \(e_2\)
  (Y, up), \(e_3\) (Z, towards viewer)
\item
  \textbf{Two null vectors:} \(n_o\) (origin) and \(n_\infty\) (point at
  infinity)
\end{itemize}

The null vectors are constructed from the extra-dimensional basis
vectors \(e_4, e_5\) as:

\[n_o = \frac{1}{2}(e_5 - e_4), \quad n_\infty = e_4 + e_5\]

satisfying \(n_o^2 = n_\infty^2 = 0\) (null) and
\(n_o \cdot n_\infty = -1\).

The motivation for adding two dimensions is that in the 5D conformal
model, rigid motions and uniform scalings are encoded in one motor
formalism, applied via the sandwich product:

\[P' = M \cdot P \cdot \widetilde{M}\]

where \(M\) is any motor (translation, rotation, dilation, or their
composition) and \(\widetilde{M}\) is its reverse.

\subsection{CGA Point Embedding}\label{32-cga-point-embedding}

A 3D Euclidean point \(\mathbf{x} = (x_1, x_2, x_3)\) is embedded into
CGA as:

\[P = n_o + x_1\,e_1 + x_2\,e_2 + x_3\,e_3 + \frac{1}{2}|\mathbf{x}|^2\,n_\infty\]

To recover the 3D point from a CGA point \(P\), we project down:

\[x_i = \frac{P \cdot e_i}{-(P \cdot n_\infty)}, \quad i \in \{1, 2, 3\}\]

This embedding is the key that enables the unified transformation
framework: once a point is in CGA form, any motor can be applied
identically through the sandwich product.

\subsection{CGA Transformation
Primitives}\label{33-cga-transformation-primitives}

All transformations in CGA are expressed as motors that encode geometric
operations.
Table~\ref{tab:1} summarizes the three fundamental motor types.

\begin{table}[htbp]
\centering
\caption{CGA transformation primitives. All are applied via the same sandwich product \(P' = M \cdot P \cdot \widetilde{M}\).}
\label{tab:1}
\begin{tabular}{@{}lll@{}}
\toprule
Motor & Formula & Description \\
\midrule
Translation \(T(\mathbf{t})\) & \(1 - \frac{1}{2}\mathbf{t}\,n_\infty\)
& Moves a point by displacement \(\mathbf{t}\) \\
Rotation \(R(\theta, u, v)\) &
\(\cos(\theta/2) - \sin(\theta/2)\,(u \wedge v)\) & Rotates by angle
\(\theta\) in the \(u\)-\(v\) plane \\
Dilation \(D(s)\) &
\(\cosh\!\bigl(\frac{\ln s}{2}\bigr) + \sinh\!\bigl(\frac{\ln s}{2}\bigr)\,(n_o \wedge n_\infty)\)
& Scales distance from origin by factor \(s\) \\
\bottomrule
\end{tabular}
\end{table}

Motor composition is achieved through the geometric product:
\(M = T \cdot R
\cdot D\) applies \(D\) first, then \(R\), then \(T\) (right-to-left
order). The critical advantage for LLM integration is that composition
is expressed as simple multiplication of compact expressions, unlike the
4×4 matrix case where each operation requires specifying 16 numerical
entries.

\subsection{CGA vs. Euclidean Matrices: Token-Level
Comparison}\label{34-cga-vs-euclidean-matrices-token-level-comparison}

Table~\ref{tab:2} compares the CGA and Euclidean matrix representations across key
properties relevant to LLM generation.

\begin{table}[htbp]
\centering
\caption{CGA vs. Euclidean 4×4 matrix representations. CGA is consistently more compact, directly impacting LLM token budgets and wall-clock latency in interactive loops.}
\label{tab:2}
\begin{tabular}{@{}lll@{}}
\toprule
Property & CGA Cl(4,1) & Euclidean 4×4 Matrices \\
\midrule
Representation & Single multivector & 4×4 = 16 numbers \\
Composition & Geometric product & Matrix multiplication \\
Rotation 90$^\circ$ around Z & \texttt{R(pi/2,\ e1,\ e2)} (6 tokens) &
\texttt{{[}[0,-1,0,0],[1,0,0,0],...{]}} (32 tokens) \\
Translation by (3,0,0) & \texttt{T(3*e1)} (4 tokens) &
\texttt{{[}[1,0,0,3],[0,1,0,0],...{]}} (32 tokens) \\
Compose T+R & \texttt{T(3*e1)*R(pi/2,e1,e2)} & 2 matrix
multiplications \\
\bottomrule
\end{tabular}
\end{table}

Fewer tokens reduce output burden and wall-clock latency in interactive
loops. Across the dedicated full-loop protocol in
Section~\ref{57-interactive-latency-and-engage-pilot}
(Table~\ref{tab:16}), lower
completion-token counts coincide with lower median latency (Compact SE3:
0.81s, Simple CGA: 1.19s, Euclidean 4×4: 1.40s). We therefore treat
token cost as an empirical predictor of runtime burden, not a
deterministic latency law.

\begin{center}\rule{0.5\linewidth}{0.5pt}\end{center}

\section{System Architecture}\label{4-system-architecture}

\subsection{Pipeline Overview}\label{41-pipeline-overview}

The system transforms natural language into 3D scene edits through four
stages:

\begin{verbatim}
NL Input → LLM / Template Engine → CGA JSON → Execute (sandwich) → Render
\end{verbatim}

For example, the instruction \emph{"Move the red sphere next to the blue
cube"} is parsed into mover/target references, emitted as
\texttt{\{"RedSphere":\ "T(2*e1)"\}}, executed via the sandwich product
\(P' = T \cdot P \cdot \widetilde{T}\), and rendered as an updated scene.
Detailed worked derivations are provided in
Section~\ref{b4-detailed-spatial-edit-derivations}.

\begin{figure}
\centering
% \pandocbounded{\includegraphics[keepaspectratio]{figures/fig1_initial_scene.png}}
% Example PNG crop (left bottom right top):
\includegraphics[width=0.7\linewidth,trim=140bp 120bp 200bp 160bp,clip]{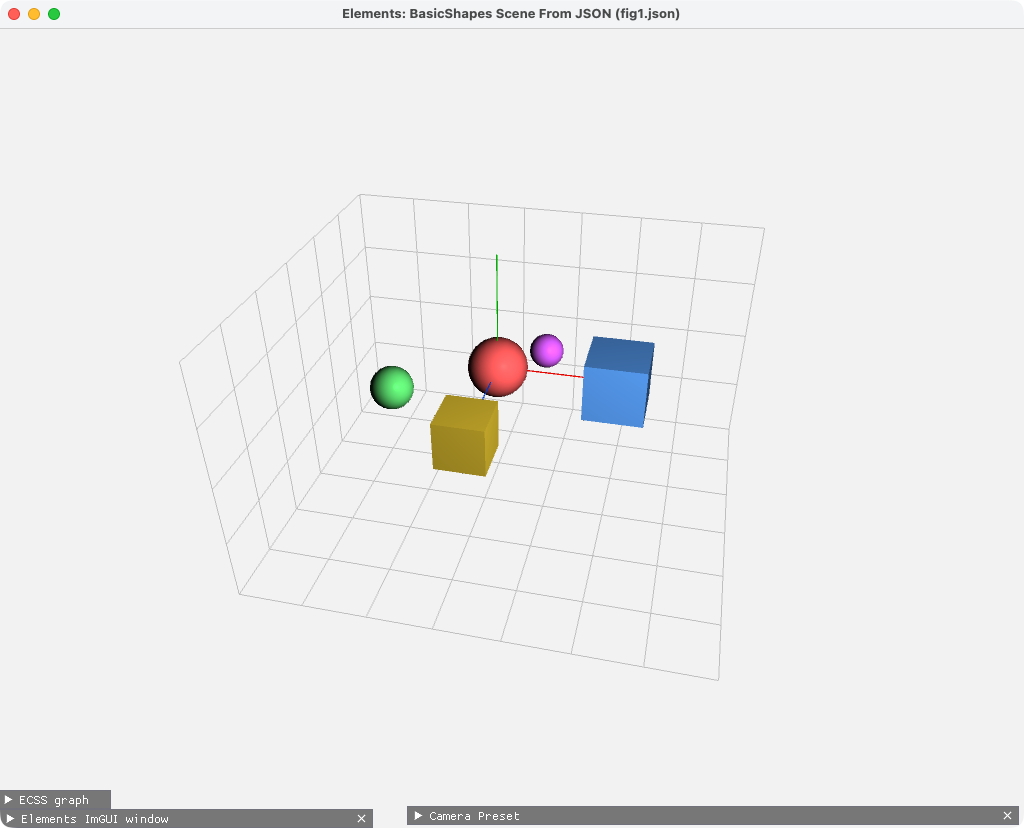}
% \pandocbounded{\includegraphics[width=0.7\linewidth]{figures/fig1.png}}
\caption{Initial 5-object test scene rendered with Y-up
convention. Objects are positioned in 3D space with CGA-compatible
coordinates: e₁ (X, right), e₂ (Y, up), e₃ (Z, towards viewer).
Visualization of the produced JSON files uses the Project Elements
framework~\cite{Elements2023}.}
\Description{Initial scene with five colored objects at fixed coordinates in a Y-up 3D frame.}
\label{fig:1}
\end{figure}

Figure~\ref{fig:1} shows the initial five-object scene used throughout
the interaction examples.

Each 3D object in the scene is stored as a \texttt{SceneObject} with
center position (Euclidean \((x,y,z)\)), bounding size (half-extent
radius), shape type (sphere or cube), and color name. The color name
serves dual purpose: display rendering and LLM reference in natural
language instructions.

The key method is \texttt{apply\_motor(motor)}, which applies a single
transformation path:
\(P = \text{cgaPoint}(x,y,z)\rightarrow P' = M \cdot P \cdot \widetilde{M}\rightarrow \text{center} = \text{down}(P')\).

This single operation is the practical advantage of CGA: one expression
path handles translations, rotations, dilations, and their compositions,
removing transformation-specific branching.

Mesh primitives are generated procedurally: spheres use an icosphere
(subdivided icosahedron with 42 vertices and 80 faces after one
subdivision), and cubes use 8 vertices with 12 triangular faces. The
coordinate convention follows standard graphics practice
(OpenGL/Blender/Unity): \(e_1\) = X (right), \(e_2\) = Y (up), \(e_3\) =
Z (towards viewer).

\subsection{Three Competing Prompting
Strategies}\label{43-three-competing-prompting-strategies}

We compare three system prompts, all using the same LLM (GPT-4o-mini),
to isolate the effect of the output representation:

\subsubsection{Simple CGA.}\label{simple-cga-ours}

A minimal prompt (\textasciitilde720 characters) specifying only the
axis conventions, available operations, and output format. The prompt
instructs the LLM to produce JSON with object names as keys and CGA
expression strings as values. A critical design choice is explicitly
requiring basis vectors in \texttt{R()} to prevent the LLM from
generating float-to-int errors (e.g., \texttt{R(pi/2,\ 1,\ 2)} instead
of the correct \texttt{R(pi/2,\ e1,\ e2)}).

\subsubsection{Shenlong CGA (Paper
~\cite{ref1}).}\label{shenlong-cga-paper-1}

A verbose chain-of-thought prompt (\textasciitilde960 characters) that
asks the LLM to follow four explicit reasoning steps: (1) document
current object positions, (2) reason about needed transformations, (3)
compute exact CGA expressions, and (4) output the final JSON. While this
mirrors best practices for complex reasoning, the additional reasoning
tokens (200--500) often push the response past the \texttt{max\_tokens}
budget, causing truncated or malformed output. Shenlong is treated in
this paper as a verbose CGA reference configuration; its lower parse
rates are partly associated with token-budget constraints rather than a
representation-level limitation (see the ablation evidence in
Section~\ref{58-ablation-diagnostics}, where the gap is linked to
prompt-length/token-budget interaction).

\subsubsection{Euclidean 4×4 Baseline.}\label{euclidean-44-baseline}

A matrix-based prompt (\textasciitilde435 characters) instructing the
LLM to produce numpy 4×4 transformation matrices. This represents the
conventional approach in existing 3D manipulation systems. Each
transformation requires specifying all 16 matrix entries, with correct
trigonometric values for rotations.

Table~\ref{tab:3} summarizes these results.

\begin{table}[htbp]
\centering
\caption{Interactive-loop efficiency profile from the powered latency protocol (\texttt{n=40} per method, GPT-4o-mini): instruction → API response → parse/execute → render-ready state. Lower tokens and lower latency are better. Latency headroom is computed relative to a 100ms interactive-loop target.}
\label{tab:3}
\begin{tabular}{@{}llll@{}}
\toprule
Method & Avg Completion Tokens & Avg Latency per Task (s) & Latency vs
100ms VR Target \\
\midrule
\textbf{Compact SE3} & \textbf{22.13} & \textbf{0.94 $\pm$ 0.28} &
\textbf{+0.84 s above target} \\
Simple CGA & 39.13 & 1.27 $\pm$ 0.38 & +1.17 s above target \\
Euclidean 4×4 & 63.15 & 2.57 $\pm$ 4.91 & +2.47 s above target \\
\bottomrule
\end{tabular}
\end{table}

Under full-loop measurement, Compact SE3 is the fastest method and
Simple CGA is second, while Euclidean 4×4 is slowest on mean latency.
All methods remain above the 100ms immersive-loop threshold, motivating
architectural optimization of the end-to-end runtime path
(Sections~\ref{57-interactive-latency-and-engage-pilot}
and~\ref{63-latency-and-interactive-viability}).

\subsection{Compact SE3 Baseline --- Full
Specification}\label{44-compact-se3-baseline--full-specification}

The Compact SE3 control baseline encodes transformations as an ordered
list (queue) of typed SE3 operations expressed in JSON. Each operation
is a typed object with a mandatory \texttt{"type"} field and
operation-specific parameters:

\begin{itemize}
\tightlist
\item
  \textbf{Translation:}
  \texttt{\{"type":\ "T",\ "v":\ {[}dx,\ dy,\ dz{]}\}} --- Euclidean
  displacement vector.
\item
  \textbf{Rotation:}
  \texttt{\{"type":\ "R",\ "axis":\ {[}ax,\ ay,\ az{]},\ "angle\_rad":\ theta\}}
  --- axis-angle representation. The axis need not be unit-normalized;
  the executor normalises it.
\item
  \textbf{Scale:} \texttt{\{"type":\ "D",\ "factor":\ s\}} --- uniform
  scale factor applied object-centrically (consistent with the CGA
  \texttt{D(s)} semantics in
  Section~\ref{46-cga-expression-executor}).
\end{itemize}

The complete output schema for a single instruction is:

\begin{verbatim}
{
  "ObjectName": [
    {"type": "T", "v": [dx, dy, dz]},
    {"type": "R", "axis": [ax, ay, az], "angle_rad": theta},
    {"type": "D", "factor": s}
  ]
}
\end{verbatim}

Operations are executed left-to-right in list order. Multi-object
instructions use multiple top-level keys. Single-operation instructions
use a one-element list. This differs from CGA motor composition, where
the geometric product applies operations right-to-left; the system
prompt for Compact SE3 explicitly specifies left-to-right execution
order to avoid ambiguity.

The system prompt for Compact SE3 is approximately 590 characters ---
similar in length to Simple CGA and shorter than Shenlong. Rotation is
encoded as axis-angle rather than quaternion or Euler angles, since
axis-angle is the most natural LLM-friendly format for expressing
rotations about a named axis (e.g., "rotate 90$^\circ$ around Z" maps directly
to \texttt{\{"axis":\ [0,0,1],\ "angle\_rad":\ 1.5708\}} without
trigonometric identity lookups).

The key difference from CGA is that Compact SE3 has no algebraic closure
property: the executor must dispatch on \texttt{"type"} and apply each
operation independently, whereas CGA motors compose algebraically into a
single motor before any execution. This distinction matters for
compositional instruction chains, as documented in the sequence-fidelity
stress test
(Section~\ref{52-sequence-fidelity-stress-result-primary-confirmatory-finding}).

\subsection{Spatial Operations}\label{45-spatial-operations}

The system supports several canonical spatial operations, each computed
by the template engine or the LLM:

\begin{itemize}
\item
  \textbf{"next to" / "left of":} The mover\textquotesingle s surface is
  placed tangent to the target\textquotesingle s surface. Computed as:
  \(\text{displacement}_x = \text{target.min}[0] - \text{mover.size} - \text{mover.center}[0]\),
  yielding \(T(d_x \cdot e_1 + d_y \cdot e_2 + d_z \cdot e_3)\).
\item
  \textbf{"on top of":} The mover\textquotesingle s bottom is placed on
  the target\textquotesingle s top surface. Computed as:
  \(\text{new\_y} = \text{target.max}[1] + \text{mover.size}\), yielding
  the corresponding translation motor.
\item
  \textbf{"between":} The mover is placed at the midpoint of two
  reference objects:
  \(\text{midpoint} = (A.\text{center} + B.\text{center}) / 2\),
  yielding \(T(\text{midpoint} - \text{mover.center})\).
\item
  \textbf{"rotate":} Direct rotation motor \(R(\theta, e_i, e_j)\) in
  the specified plane.
\item
  \textbf{"scale":} Dilation motor \(D(s)\) applied to the
  object\textquotesingle s size (center preserved in the execution
  engine, equivalent to uniform object-centric scaling for this
  benchmark).
\end{itemize}

\subsection{CGA Expression
Executor}\label{46-cga-expression-executor}

The \texttt{execute\_cga()} function processes each generated expression
through four steps:

\begin{enumerate}
\def\labelenumi{\arabic{enumi}.}
\tightlist
\item
  \textbf{Parse} the expression string (e.g.,
  \texttt{"T(3*e1)*R(np.pi/2,e1,e2)"}).
\item
  \textbf{Evaluate} in a sandboxed environment containing only CGA
  primitives (\(e_1, e_2, e_3, n_o, n_\infty, T, R, D\)) --- no general
  Python execution.
\item
  \textbf{Apply} the motor: \texttt{obj.apply\_motor(motor)} →
  \(P' = M \cdot P \cdot
  \widetilde{M}\).
\item
  \textbf{Special case:} Pure dilation \(D(s)\) scales \texttt{obj.size}
  directly while keeping the center fixed. This realizes the
  benchmark\textquotesingle s "scale" command as object-centric uniform
  scaling rather than world-origin dilation. The Compact SE3
  \texttt{"D"} operation uses the same object-centric semantics,
  ensuring a fair comparison on scale tasks across both representations.
\end{enumerate}

\subsection{Engine Selection Logic}\label{47-engine-selection-logic}

The pipeline uses a hybrid approach for robust operation:

\begin{itemize}
\tightlist
\item
  \textbf{Known spatial keywords} (e.g., "on top", "next to", "between",
  "rotate", "scale") → deterministic template engine
  (\texttt{template\_cga}), guaranteeing correct results.
\item
  \textbf{Novel instructions} + LLM available → \texttt{call\_llm}
  (GPT-4o-mini with Simple CGA prompt).
\item
  \textbf{No API key} → template fallback for graceful degradation.
\end{itemize}

This design ensures the system is fully functional without API access
while still enabling open-ended language understanding when the LLM is
available. For benchmarking clarity, Section~\ref{5-experiments-and-results} does not mix these
routes: reported benchmark metrics are collected with direct
model-generation harnesses, while template routing is used for
interactive scene-edit sessions.

\subsection{Natural Language Scene Editing
Cases}\label{48-natural-language-scene-editing-cases}

\begin{figure}
\centering
% \pandocbounded{\includegraphics[keepaspectratio]{figures/fig2_side_placement_before_after.png}}
\includegraphics[width=0.49\linewidth,trim=140bp 120bp 200bp 160bp,clip]{figures/fig1.png}
\includegraphics[width=0.49\linewidth,trim=140bp 120bp 200bp 160bp,clip]{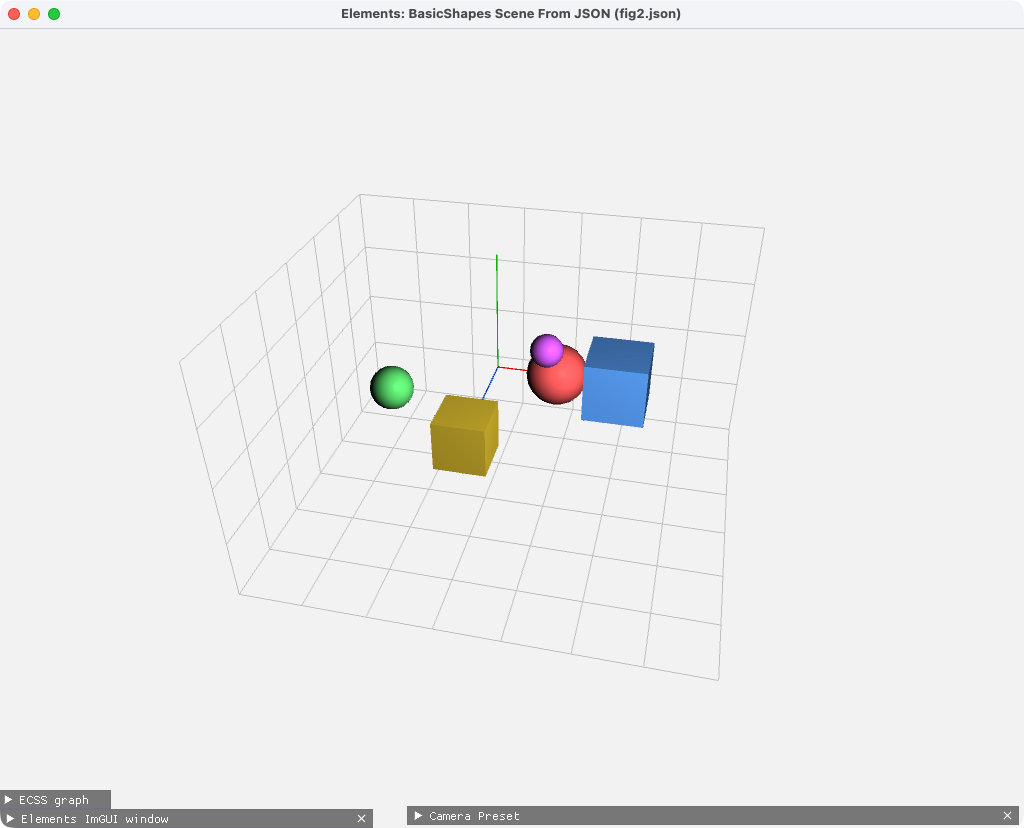} % plus initial
\caption{Before/after visualization of a side-placement edit.
The CGA translation motor \(T(2 \cdot e_1)\) moves RedSphere from the origin to
x=2, placing its surface tangent to BlueCube.}
\Description{Two panels show the red sphere moved left of the blue cube so their surfaces are tangent.}
\label{fig:2}
\end{figure}

Figure~\ref{fig:2} shows the before/after result of the instruction \emph{"Move
the red sphere next to the blue cube, to its left side."} The template
engine computes the displacement
\(d_x = \text{target.min}[0] - \text{mover.size} -
\text{mover.center}[0] = 3 - 1.0 - 0 = 2.0\), generates
\texttt{T(2.0*e1)}, and applies the sandwich product. The red sphere
moves from [0,0,0] to [2,0,0], with its right surface touching
the blue cube\textquotesingle s left surface at \(x=3\).

\begin{figure}
\centering
% \pandocbounded{\includegraphics[keepaspectratio]{figures/fig3_vertical_stacking.png}}
\includegraphics[width=0.49\linewidth,trim=140bp 120bp 200bp 160bp,clip]{figures/fig2.png}
\includegraphics[width=0.49\linewidth,trim=140bp 120bp 200bp 160bp,clip]{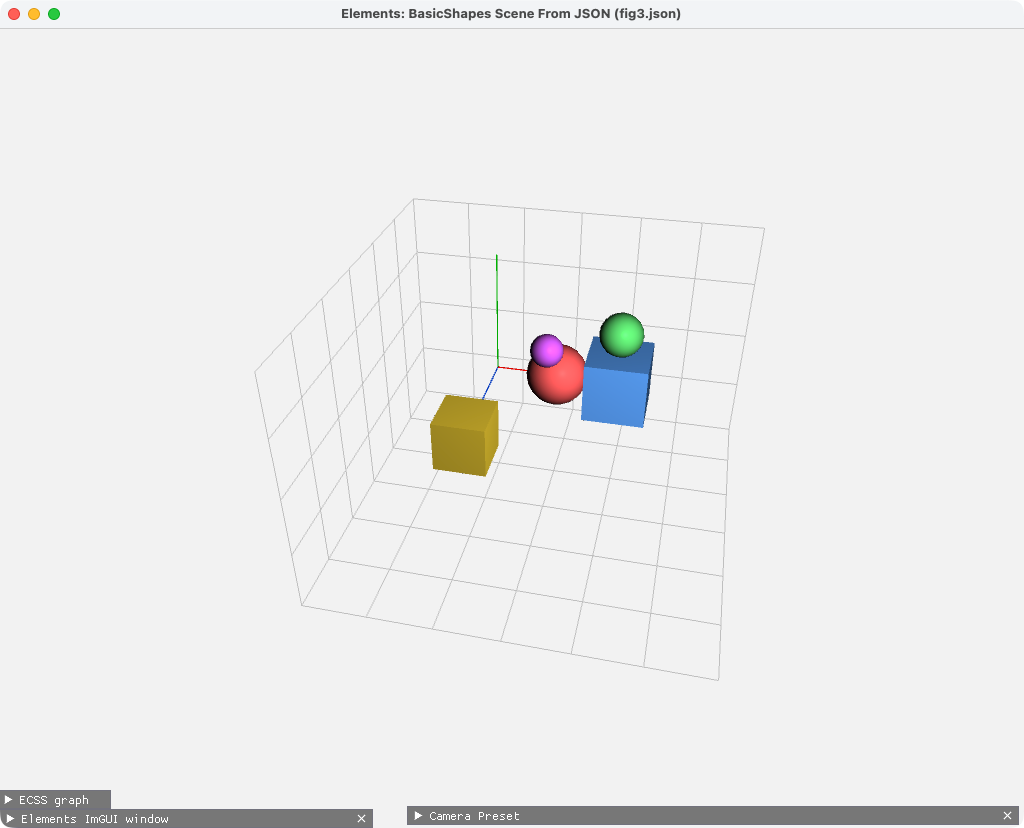}
\caption{Vertical stacking case. The CGA motor translates
GreenSphere to sit precisely on top of BlueCube.}
\Description{Two panels show the green sphere translated to rest exactly on top of the blue cube.}
\label{fig:3}
\end{figure}

Figure~\ref{fig:3} demonstrates vertical stacking: \emph{"Place the green sphere
on top of the blue cube."} The engine computes
\texttt{new\_y\ =\ BlueCube.max[1]\ +\ GreenSphere.size\ =\ 1.0\ +\ 0.7\ =\ 1.7}
and generates the displacement \texttt{T(7.0*e1\ +\ 1.7*e2\ −\ 2.0*e3)},
moving GreenSphere from {[}−3,0,2{]} to {[}4,1.7,0{]}. Verification:
GreenSphere bottom = 1.7 − 0.7 = 1.0 = BlueCube top.

\begin{figure}
\centering
% \pandocbounded{\includegraphics[keepaspectratio]{figures/fig4_sequential_editing.png}}
\includegraphics[width=0.24\linewidth,trim=140bp 120bp 200bp 160bp,clip]{figures/fig1.png}
\includegraphics[width=0.24\linewidth,trim=140bp 120bp 200bp 160bp,clip]{figures/fig2.png}
\includegraphics[width=0.24\linewidth,trim=140bp 120bp 200bp 160bp,clip]{figures/fig3.png}
\includegraphics[width=0.24\linewidth,trim=140bp 120bp 200bp 160bp,clip]{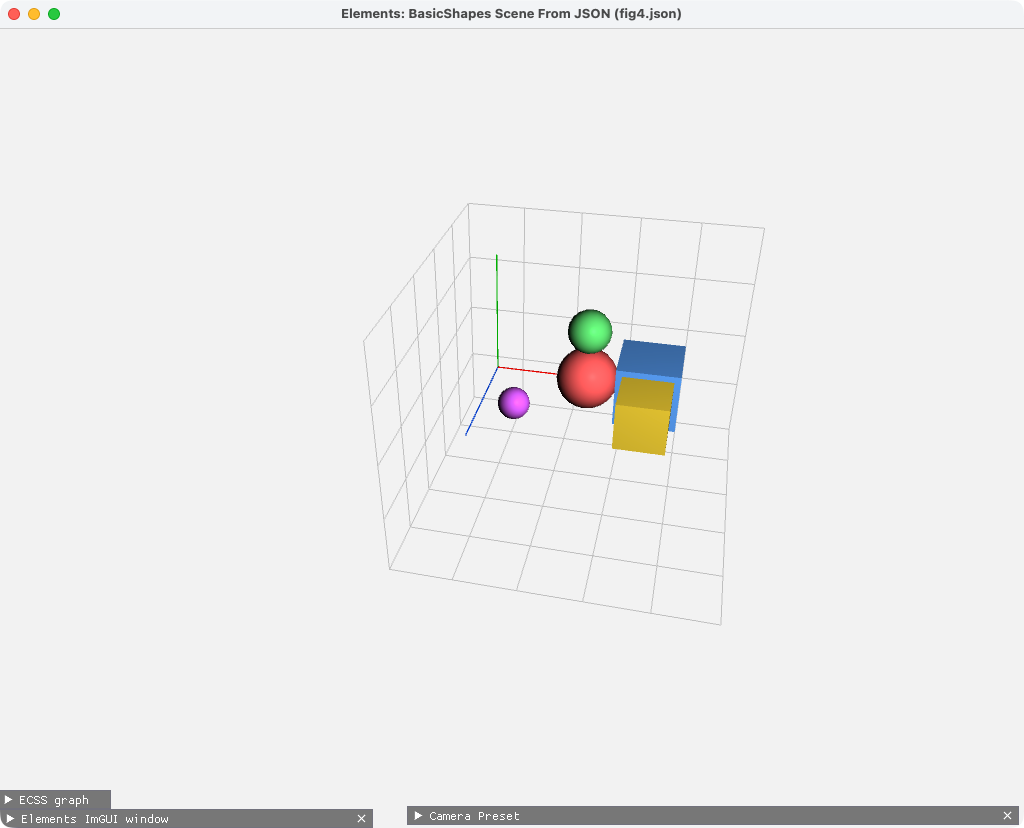}
\caption{Sequential scene editing pipeline. Three natural language
instructions are applied in sequence: "Move the red sphere next to the
blue cube on its left.", "Stack the green sphere on top of the red
sphere.", and "Move the yellow cube in front of the blue cube." Each
step builds on the preceding scene state.}
\Description{Four-step pipeline showing the initial scene and three successive edits after each instruction.}
\label{fig:4}
\end{figure}

Figure~\ref{fig:4} shows sequential editing --- three consecutive natural language
instructions building a scene arrangement. Each instruction is applied
to the scene state left by the preceding one, demonstrating
CGA\textquotesingle s composability across operations.

\begin{figure}
\centering
\pandocbounded{\includegraphics[keepaspectratio]{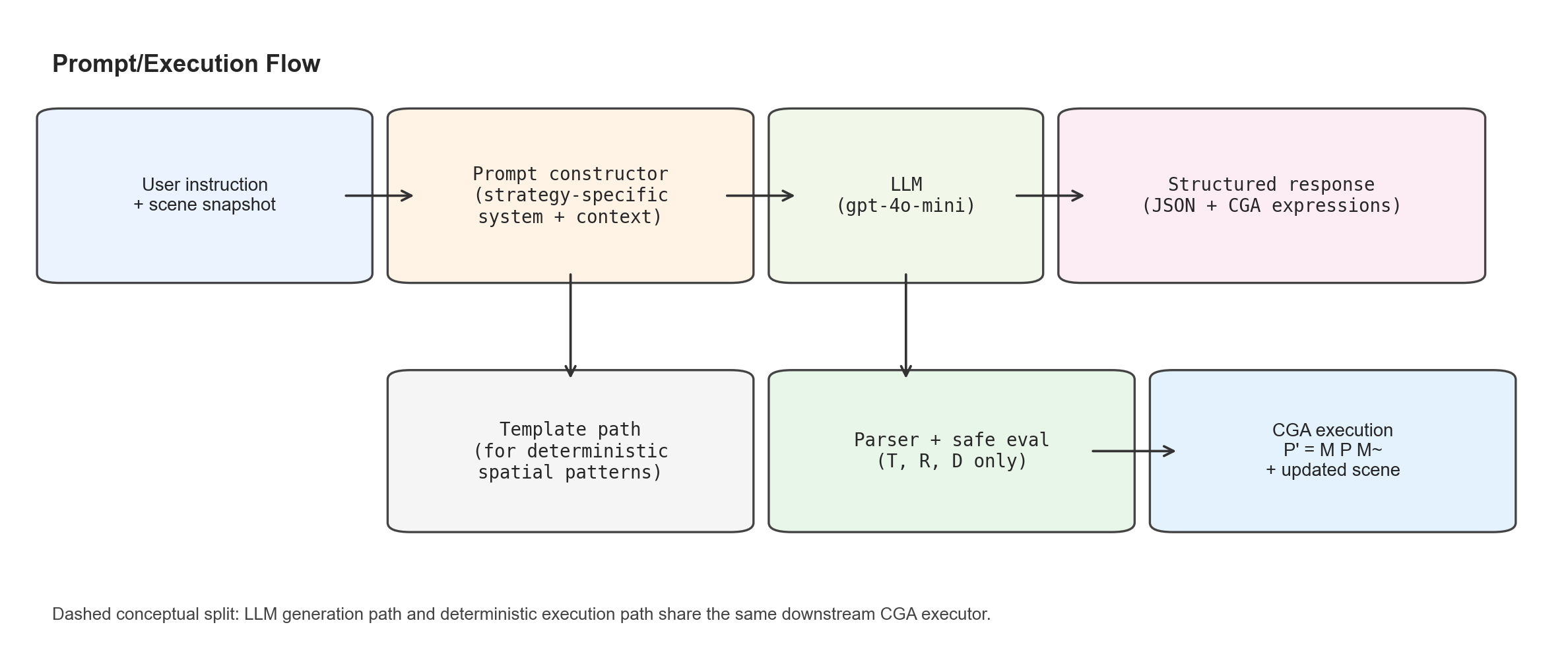}}
\caption{Prompt-to-execution flow used in all benchmarks.}
\Description{Flow diagram from natural-language prompt through parsing, engine routing, execution, and rendering.}
\label{fig:5}
\end{figure}

Figure~\ref{fig:5} visualizes the full execution path, including the
template-assisted branch and the shared safe-execution backend.

The template-assisted branch bypasses model generation but converges to
the same executor.

\begin{figure}
\centering
\pandocbounded{\includegraphics[keepaspectratio]{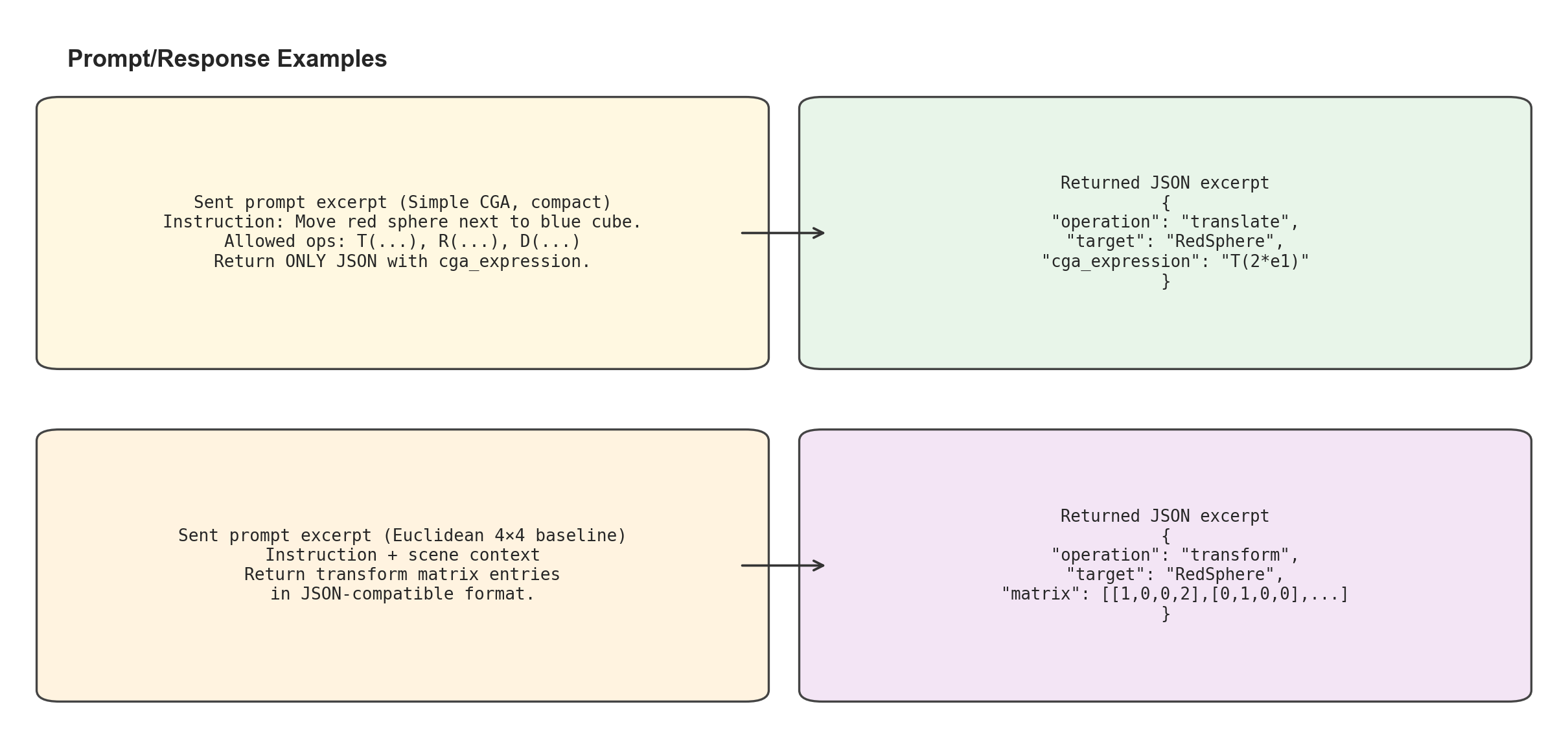}}
\caption{Prompt/response examples for CGA and Euclidean
baselines.}
\Description{Example prompt and model outputs for compact CGA and Euclidean matrix representations.}
\label{fig:6}
\end{figure}

Figure~\ref{fig:6} provides representative prompt/response snippets for the
compact CGA and Euclidean baselines, clarifying what is sent to the
model and what is returned for execution.

\begin{center}\rule{0.5\linewidth}{0.5pt}\end{center}

\section{Experiments and Results}\label{5-experiments-and-results}

This section is organized by \textbf{evidential strength}, leading with
the result that carries the strongest statistical support. \texttt{Q1}
(sequence fidelity): does algebraic expression form affect ordered-chain
faithfulness beyond compactness alone? \texttt{Q2} (reliability at
scale): does representation choice change success and cost as scene
complexity grows? \texttt{Q3} (stability): are observed gains robust
under repeated runs? \texttt{Q4} (semantic validity): do parse-valid
outputs remain geometrically correct under stricter checks? \texttt{Q5}
(mechanism): how much of the observed behavior is attributable to
representation, prompt style, and retry policy?

Q1 is addressed in
Section~\ref{52-sequence-fidelity-stress-result-primary-confirmatory-finding}
(the confirmed primary result), Q2 in
Section~\ref{53-supporting-reliability-benchmark}, Q3 in
Section~\ref{54-stability-under-repeated-runs}, Q4 in
Section~\ref{55-semantic-validity-under-harder-grounding}, and Q5 in
Section~\ref{58-ablation-diagnostics}.
Section~\ref{56-multi-model-replication} consolidates multi-model
replication, and Section~\ref{57-interactive-latency-and-engage-pilot}
reports interactive latency in an immersive editing loop. Following
common LLM-evaluation practice, each block pairs compact summary tables
with trend-focused figures, while exhaustive rows remain in Appendix A.

\subsection{Evaluation Protocol and
Terminology}\label{51-evaluation-protocol-and-terminology}

All experiments use the same model family (\texttt{gpt-4o-mini}). Core
benchmark blocks compare three interface strategies: Shenlong CGA,
Simple CGA, and the Euclidean 4×4 baseline;
Sections~\ref{52-sequence-fidelity-stress-result-primary-confirmatory-finding}
and~\ref{55-semantic-validity-under-harder-grounding} add a
Compact SE3 control baseline. The benchmarks execute concrete
scene-edit tasks rather than synthetic symbolic strings, with outcomes
measured at execution time (parse/evaluate validity), geometric error
time (spatial accuracy), and semantic satisfaction time (geometry-based
task checks).

Terminology used in this section:

\begin{enumerate}
\def\labelenumi{\arabic{enumi}.}
\tightlist
\item
  \texttt{compact} prompt: the Simple CGA variant that encourages
  minimal symbolic output, e.g., \texttt{T(2*e1)} for a +2 x-axis
  translation.
\item
  \texttt{verbose} prompt: a more explicit Simple CGA variant that tends
  to emit expanded forms, e.g., \texttt{T(2.0*e1\ +\ 0.0*e2\ +\ 0.0*e3)}
  for the same operation.
\item
  \texttt{policy}: the per-task retry budget and decoding schedule.
  \texttt{pass@1} means one attempt; \texttt{pass@2} means up to two
  attempts; stress tests report \texttt{pass@3}.
\end{enumerate}

Pre-specified inferential protocol (six items, fixed prior to data
collection):

\begin{enumerate}
\def\labelenumi{\arabic{enumi}.}
\tightlist
\item
  Primary endpoint: sequence-fidelity under compositional stress
  (\texttt{n=120} per method); and semantic success on hard grounding
  suites (\texttt{n=20}, \texttt{n=100}).
\item
  Primary contrasts: Simple CGA vs Compact SE3 (sequence-fidelity);
  Simple CGA vs Euclidean 4×4, and Compact SE3 vs Euclidean 4×4
  (semantic suite).
\item
  Secondary contrasts: Shenlong vs Euclidean, and Simple CGA vs Compact
  SE3 (semantic suite).
\item
  Test family: two-sided Fisher exact tests on success counts;
  two-proportion z-test on sequence-fidelity rates.
\item
  Effect-size reporting: risk difference (pp) and relative risk, each
  with 95\% confidence intervals, reported alongside p-values.
\item
  P-value interpretation: unless otherwise stated, inference uses a
  two-sided \(\alpha=0.05\) threshold; \texttt{p=1.0} indicates no
  detectable difference in these data (not proof of equivalence).
\end{enumerate}

To remove template-routing confounding in evaluation, benchmark runs in
Section~\ref{5-experiments-and-results} are executed through direct model-generation protocol paths
(no keyword template dispatch during metric collection). Route
accounting is summarized below.

\textbf{Statistical power planning.} For the primary semantic contrast
(compact vs Euclidean, +20 pp at baseline 25\%), the minimum sample size
for confirmatory inference is determined by the power calculation in
Table~\ref{tab:4}. The powered suite (\texttt{n=100} per arm) crosses the 80\%
threshold and is used for confirmatory semantic inference in
Section~\ref{55-semantic-validity-under-harder-grounding}.

\begin{table}[htbp]
\centering
\caption{Statistical power calculation for primary semantic contrast (two-sided Fisher exact, Simple CGA vs Euclidean 4×4, baseline rate 25\%, expected rate 45\%). \texttt{n=80} is the minimum target for confirmatory status; the powered suite reports achieved power at \texttt{n=100}.}
\label{tab:4}
\begin{tabular}{@{}lllllll@{}}
\toprule
Contrast & Baseline Rate & Expected Rate & Effect Size & α & Power
Target & Required n per Arm \\
\midrule
Compact vs Euclidean & 25\% & 45\% & +20 pp & 0.05 & 0.80 &
\textasciitilde80 \\
Compact vs Euclidean & 25\% & 45\% & +20 pp & 0.05 & 0.90 &
\textasciitilde107 \\
Powered semantic suite & 24\% & 45\% & +21 pp & 0.05 & actual &
\textasciitilde88\% (n=100) \\
\bottomrule
\end{tabular}
\end{table}

The sequence-fidelity contrast
(Section~\ref{52-sequence-fidelity-stress-result-primary-confirmatory-finding})
is powered by design:
\texttt{n=120} per method under a +7.5 pp effect at 90\% baseline yields
\textgreater95\% power; together with the powered semantic suite, this
provides two confirmatory result blocks. Routing disclosure and
route-split metrics are reported in
Tables~\ref{tab:5} and~\ref{tab:5b}.

Table~\ref{tab:5} summarizes these results.

\begin{table}[htbp]
\centering
\caption{Routing disclosure summary (Section~\ref{5-experiments-and-results} benchmark protocol).}
\label{tab:5}
\begin{tabular}{@{}llll@{}}
\toprule
Benchmark block & Tasks / rows & Template-routed & Pure LLM-routed \\
\midrule
\makecell[l]{Core benchmark\\(5-object + stress + 10-object + spatial + 100-object)} &
33 tasks & 0 (0.0\%) & 33 (100.0\%) \\
Expanded exact-placement suite & 18 tasks & 0 (0.0\%) & 18 (100.0\%) \\
Semantic subset & 19 tasks & 0 (0.0\%) & 19 (100.0\%) \\
Hard-pack & 20 tasks & 0 (0.0\%) & 20 (100.0\%) \\
Powered hard semantic suite & 100 tasks per method & 0 (0.0\%) & 100
(100.0\%) \\
Repeated ablation block & 400 evaluated rows & 0 (0.0\%) & 400
(100.0\%) \\
\bottomrule
\end{tabular}
\end{table}

\begin{table}[htbp]
\centering
\caption{Route-split metric summary (all benchmark rows are LLM-route rows).}
\label{tab:5b}
\begin{tabular}{@{}lllll@{}}
\toprule
Benchmark metric (\texttt{route\ =\ LLM}) & Shenlong CGA & Simple CGA &
Euclidean 4×4 & Compact SE3 \\
\midrule
Semantic subset semantic success (\texttt{n=19}) & 9/19 & \textbf{10/19}
& 4/19 & - \\
Hard-pack semantic success (\texttt{n=20}) & \textbf{9/20} &
\textbf{9/20} & 5/20 & \textbf{9/20} \\
Powered hard semantic success (\texttt{n=100}) & 44/100 &
\textbf{45/100} & 24/100 & 42/100 \\
\bottomrule
\end{tabular}
\end{table}

The executed core blocks are: 5-object scene (\texttt{n=8}),
compositional stress tasks (\texttt{n=6}), 10-object scene
(\texttt{n=6}), and 100-object scene (\texttt{n=10}). For exact
placement, we report an expanded spatial-accuracy suite (\texttt{n=18})
in place of the earlier diagnostic \texttt{n=3} probe; these results are
used inferentially in Section~\ref{53-supporting-reliability-benchmark}
and Appendix A (Table~\ref{tab:a4}). Success
metrics are block-specific: parse/evaluate for general edits, pass@3 for
stress tasks, error \texttt{\textless{}\ 0.5} for exact placements, and
semantic geometry checks for semantic subsets. Because these blocks use
different success criteria, we report them separately and do not treat
any core-wide total as a single inferential metric.

Protocol metadata (token limits, retry settings) is
provided in Appendix A (Table~\ref{tab:a1}).

\subsection{Sequence-Fidelity Stress Result (Primary Confirmatory
Finding)}\label{52-sequence-fidelity-stress-result-primary-confirmatory-finding}

To isolate the contribution of algebraic expression form beyond
compactness, we evaluate strict operation-sequence fidelity under
compositional stress in a compact-vs-compact design. The protocol uses
20 sequence templates with 6 trials each, yielding \texttt{n=120}
outputs per method. Templates contain ordered chains of 3--5 distinct
spatial operations (translate, rotate, scale, and compositions thereof)
applied to named objects. The evaluation criterion is \emph{exact}
ordered-chain preservation: all operations in the specified sequence
must appear in the correct order in the model\textquotesingle s output.
Parse validity (correct JSON + evaluable CGA/SE3 expressions) is a
necessary but not sufficient condition for sequence-fidelity success.

Both Simple CGA and Compact SE3 reach parse saturation (120/120, 100\%),
ruling out any parse-rate confound in the fidelity comparison. Under
this parse-saturated condition, Simple CGA preserves exact ordered
operation chains more reliably (117/120, 97.5\%) than Compact SE3
(108/120, 90.0\%), while also using fewer completion tokens on
parse-success rows (112.6 vs 133.6).

Table~\ref{tab:6} summarizes these results.

\begin{table}[htbp]
\centering
\caption{Sequence-fidelity stress check (compact-vs-compact): strict ordered-chain preservation under \texttt{n=120} outputs per method (20 templates × 6 trials). Higher fidelity is better; lower tokens are better. This is the primary confirmatory finding of the paper.}
\label{tab:6}
\small
\begin{tabular}{@{}p{0.32\linewidth}p{0.30\linewidth}p{0.30\linewidth}@{}}
\toprule
Metric & Simple CGA & Compact SE3 \\
\midrule
Parse success & 120/120 & 120/120 \\
Parse rate (Wilson 95\% CI) & \makecell[l]{100.0\%\\{[}96.9\%, 100.0\%{]}} &
\makecell[l]{100.0\%\\{[}96.9\%, 100.0\%{]}} \\
Sequence-fidelity success & \textbf{117/120} & 108/120 \\
Sequence-fidelity rate (Wilson 95\% CI) &
\makecell[l]{\textbf{97.5\%}\\\textbf{{[}92.9\%, 99.1\%{]}}} &
\makecell[l]{90.0\%\\{[}83.3\%, 94.2\%{]}} \\
Avg tokens (parse-success rows) & \textbf{112.6} & 133.6 \\
\bottomrule
\end{tabular}
\end{table}

The pairwise sequence-fidelity difference is +7.5 pp in favor of Simple
CGA (95\% CI {[}1.4, 13.6{]}, two-proportion \texttt{p=0.016}). The
21.0-token gap indicates lower generation burden for Simple CGA on this
endpoint; end-to-end latency is evaluated directly and separately in
Section~\ref{57-interactive-latency-and-engage-pilot}.

Because outputs are clustered by template (20 templates × 6 trials), the
row-level z-test can overstate effective sample size. As a sensitivity
check, we compared per-template fidelity rates across the same 20
templates. The directional difference remains (+7.5 pp mean), but 17/20
templates are ties and a two-sided sign test on non-tied templates is
inconclusive (\texttt{n=3}, \texttt{p=1.0}).

A plausible mechanistic account is that CGA\textquotesingle s motor
formalism expresses a composed operation chain as a single geometric
product --- \texttt{T(v)*R(angle,u,v)*D(s)} is one algebraic object with
a defined and unambiguous execution order. The Compact SE3
representation encodes the same chain as an ordered list of typed JSON
objects, where the ordering must be inferred from list position. Under
adversarial or ambiguous instruction phrasings, the list-order encoding
appears more susceptible to reordering than the algebraic product
encoding. This result is consistent with the hypothesis that algebraic
expression form --- independently of compactness --- supports
compositional faithfulness in LLM-to-geometry pipelines.

\subsection{Supporting Reliability
Benchmark}\label{53-supporting-reliability-benchmark}

To test whether representation effects persist beyond one scene size, we
report the operational core blocks together with an expanded
exact-placement suite (\texttt{n=18}). The condition-level summary is
reported in Table~\ref{tab:7}. As disclosed in Section~\ref{51-evaluation-protocol-and-terminology}, all rows are LLM-route
benchmark rows (template-routed rows: 0).

\begin{table}[htbp]
\centering
\caption{Core benchmark summary across all operational regimes. Values report task success and, where applicable, average completion tokens. Higher success is better. For tokens, lower is better. For spatial accuracy, lower Euclidean error is better and success denotes placements with error \texttt{\textless{}\ 0.5}. Spatial-accuracy rates include Wilson 95\% CIs, reported as {[}lower, upper{]}.}
\label{tab:7}
\footnotesize
\begingroup
\setlength{\tabcolsep}{1pt}
\begin{tabular}{@{}p{0.11\linewidth}p{0.20\linewidth}p{0.15\linewidth}p{0.15\linewidth}p{0.15\linewidth}p{0.18\linewidth}@{}}
\toprule
Condition & What was tested & Shenlong CGA & Simple CGA & Euclidean 4×4
& Key observation \\
\midrule
\makecell[l]{5-object\\(\texttt{n=8})} & \makecell[l]{Translate, stack, scale, rotate,\\compose, multi-object,\\and hard spatial edits} & 6/8 (75\%), 51 tok & \makecell[l]{\textbf{8/8 (100\%),}\\\textbf{33 tok}} & \makecell[l]{7/8 (88\%),\\58 tok} & Simple CGA is best on both
reliability and cost. \\
\makecell[l]{Stress\\(\texttt{n=6})} & \makecell[l]{Irrational-angle\\rotation, triple\\composition, chained\\rotation, global\\scaling, relative\\stacking} & \textbf{6/6} &
\textbf{6/6} & 5/6 & \makecell[l]{Both CGA variants\\saturate this\\compositional set.} \\
\makecell[l]{10-object\\(\texttt{n=6})} & \makecell[l]{Scene edits with larger\\context and multi-object\\references} & \textbf{6/6, 46 tok} & 6/6, 57 tok & \makecell[l]{6/6,\\101 tok} & Success ties; token cost differs strongly. \\
\makecell[l]{Spatial accuracy\\(\texttt{n=18})} & \makecell[l]{Expanded exact target\\placements (absolute\\coordinates + displacement\\constraints)} & \makecell[l]{\textbf{18/18}\\\textbf{100.0\%}\\\textbf{{[}82.4\%, 100.0\%{]}}\\\textbf{err 0.00}} & \makecell[l]{\textbf{18/18}\\\textbf{100.0\%}\\\textbf{{[}82.4\%, 100.0\%{]}}\\\textbf{err 0.00}} & \makecell[l]{5/18\\27.8\%\\{[}12.5\%, 50.9\%{]}\\err 2.35} & Both CGA variants saturate exact-placement success; Euclidean
is substantially lower with larger geometric error. \\
\makecell[l]{100-object\\(\texttt{n=10},\\30-context)} & \makecell[l]{Edits in procedurally\\generated large scene} & 9/10 (90\%), 64 tok & \textbf{10/10 (100\%), 47 tok} &
10/10 (100\%), 86 tok & Simple and Euclidean tie on success; Simple is
much cheaper. \\
\bottomrule
\end{tabular}
\endgroup
\end{table}

To preserve readability, per-task detail tables and supplementary
condition figures are moved to Appendix A
(Tables~\ref{tab:a2}--\ref{tab:a5},
Figures~\ref{fig:a1}--\ref{fig:a3}).

For the expanded spatial suite, pairwise exact-success tests (two-sided
Fisher) show Euclidean vs Shenlong: \texttt{p\textless{}1e-5}, and
Euclidean vs Simple CGA: \texttt{p\textless{}1e-5}; Shenlong vs Simple
CGA is identical (\texttt{p=1.0}).

\begin{figure}
\centering
\pandocbounded{\includegraphics[keepaspectratio]{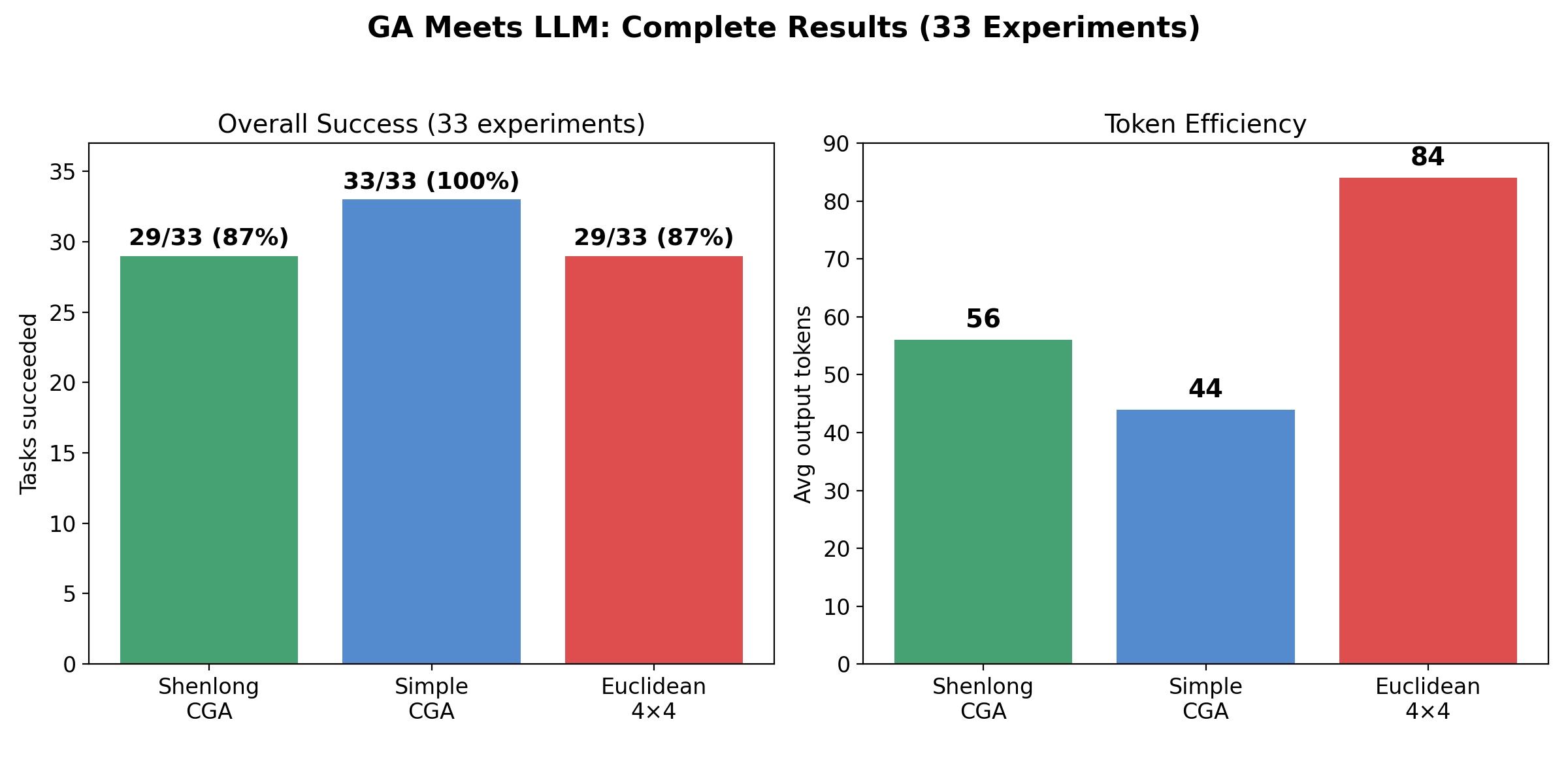}}
\caption{Core-block mixed-endpoint overview across evaluated
conditions.}
\Description{Summary chart of benchmark outcomes across methods and evaluation conditions.}
\label{fig:7}
\end{figure}

Figure~\ref{fig:7} summarizes the core-block mixed-endpoint outcomes
across all evaluated conditions.

\begin{figure}
\centering
\pandocbounded{\includegraphics[keepaspectratio]{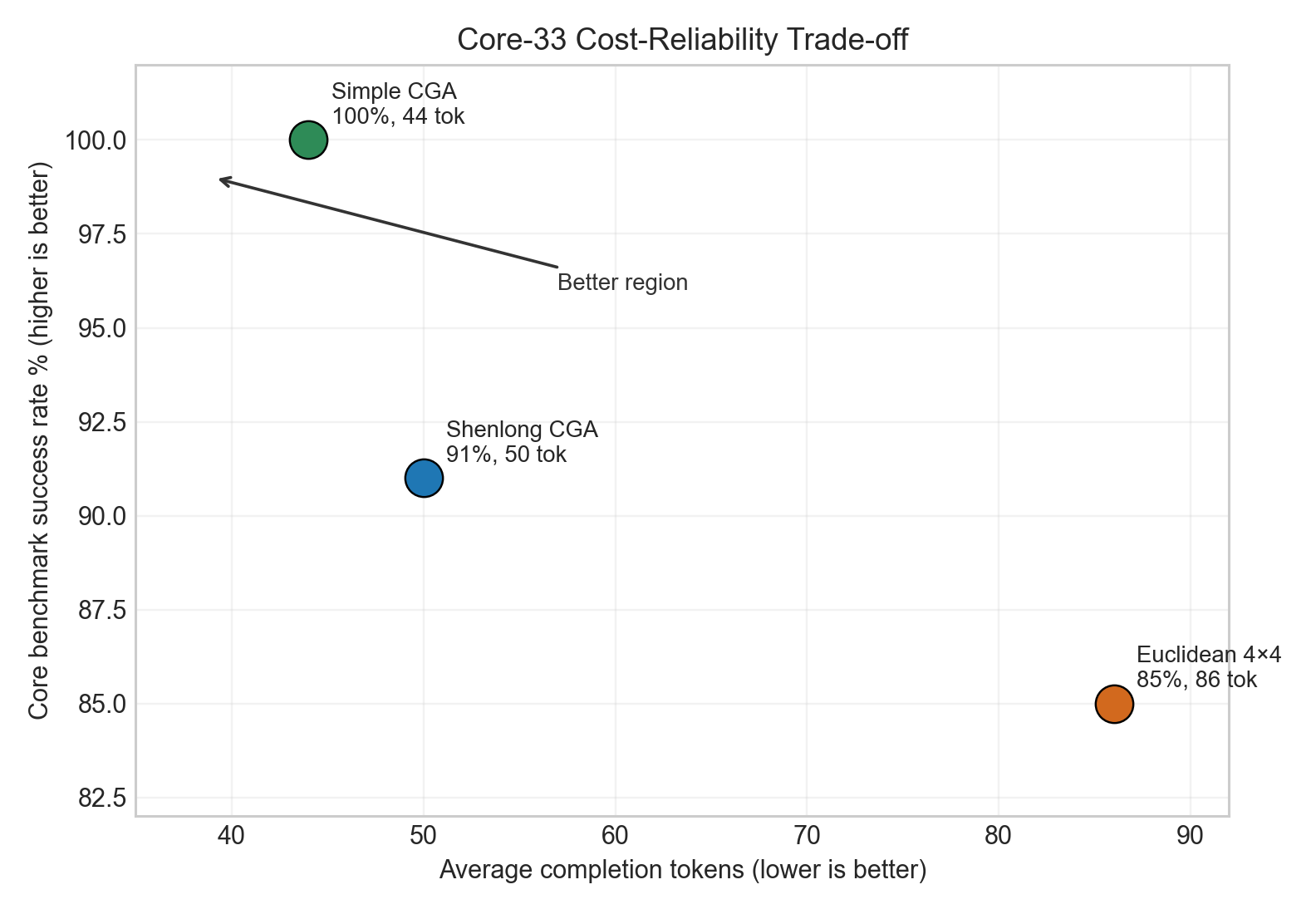}}
\caption{Core-block mixed-endpoint success/token Pareto view.}
\Description{Pareto-style plot comparing task success against completion token cost across methods.}
\label{fig:8}
\end{figure}

To make the trade-off explicit in one consolidated view, Figure~\ref{fig:8}
plots an operational mixed-endpoint Pareto plane (success vs. completion
tokens), while formal claims remain tied to endpoint-specific rows and
hard-suite semantics.

This plot is included as a descriptive operational view rather than an
inferential result.

\subsection{Stability Under Repeated
Runs}\label{54-stability-under-repeated-runs}

\begin{figure}
\centering
\pandocbounded{\includegraphics[keepaspectratio,trim=0bp 0bp 0bp 36bp,clip]{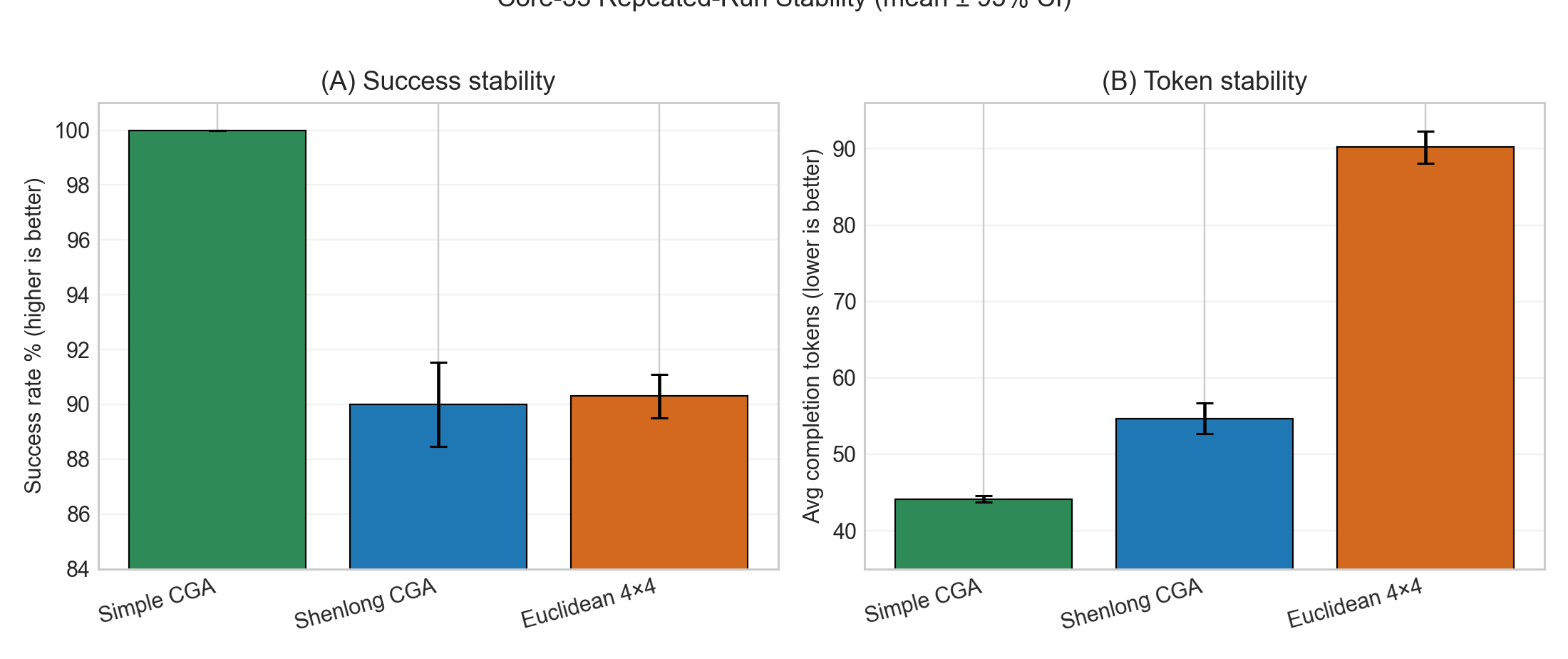}}
\caption{Core-33 repeated-run stability as mean $\pm$ 95\% CI.}
\Description{Repeated-run stability plot with mean values and 95 percent confidence intervals for each method.}
\label{fig:9}
\end{figure}

To separate stable effects from single-run variance, we repeat the full
33-task benchmark 10 times under fixed prompts and fixed default policy
settings. Table~\ref{tab:8} reports success and token uncertainty statistics, and
Figure~\ref{fig:9} visualizes the same means and confidence bands.

\begin{table}[htbp]
\centering
\caption{Core-33 uncertainty summary across repeated runs (\texttt{n=10} runs). Higher success mean is better. Lower SD/CI indicates higher stability. For token metrics (tok), lower values are better.}
\label{tab:8}
\small
\begin{tabular}{@{}lcccccc@{}}
\toprule
Approach & \makecell{Success\\Mean} & \makecell{Success\\SD} &
\makecell{Success\\95\% CI} & \makecell{Avg Tokens\\Mean} &
\makecell{Avg Tokens\\SD} & \makecell{Avg Tokens\\95\% CI} \\
\midrule
Shenlong CGA & 29.70 & 0.82 & +/-0.51 & 54.70 & 3.24 & +/-2.01 \\
\textbf{Simple CGA} & \textbf{33.00} & \textbf{0.00} & \textbf{+/-0.00}
& \textbf{44.13} & \textbf{0.69} & \textbf{+/-0.43} \\
Euclidean 4×4 & 29.80 & 0.42 & +/-0.26 & 90.18 & 3.41 & +/-2.11 \\
\bottomrule
\end{tabular}
\end{table}

The zero-SD result for Simple CGA is notable: across 10 independent runs,
this method achieves identical task success with near-zero token
variance. This pattern is consistent with higher run-to-run stability for
the Simple CGA configuration in this benchmark.

Figure~\ref{fig:9} is consistent with a strong and stable Simple CGA
profile in this protocol.

\subsection{Semantic Validity Under Harder
Grounding}\label{55-semantic-validity-under-harder-grounding}

Because parse-valid outputs can still diverge from geometric intent in
harder settings, semantic checks are evaluated explicitly on two
datasets: a semantic subset (\texttt{n=19}) and a harder
language-grounding pack (\texttt{n=20}) with paraphrase, distractor,
noisy-phrasing, and longer compositional instructions. Semantic outcomes
are validated by geometry rules (surface contact, midpoint constraints,
target displacements, scale factors, and absolute placement checks).

\textbf{Calibration note:} This section includes descriptive subsets
(\texttt{n=19}, \texttt{n=20}) and the powered semantic suite
(\texttt{n=100} per method). The powered suite is used for confirmatory
semantic inference, while smaller blocks provide descriptive support.

Table~\ref{tab:9} summarizes these results.

\begin{table}[htbp]
\centering
\caption{Parse-validity versus semantic-validity outcomes. Parse columns capture executable output structure; semantic columns capture geometric task satisfaction. Higher rates are better. For tokens, lower is better.}
\label{tab:9}
\small
\begin{tabular}{@{}p{0.17\linewidth}lcccc>{\centering\arraybackslash}p{0.20\linewidth}@{}}
\toprule
Dataset & Approach & \makecell{Parse\\Success} & \makecell{Parse\\Rate}
& \makecell{Semantic\\Success} & \makecell{Semantic\\Rate} &
\makecell{Avg Tokens\\(parse-success rows)} \\
\midrule
Semantic subset (\texttt{n=19}) & Shenlong CGA & \textbf{19/19} &
\textbf{100.0\%} & 9/19 & 47.4\% & 60.42 \\
Semantic subset (\texttt{n=19}) & Simple CGA & \textbf{19/19} &
\textbf{100.0\%} & \textbf{10/19} & \textbf{52.6\%} & \textbf{41.37} \\
Semantic subset (\texttt{n=19}) & Euclidean 4×4 & 18/19 & 94.7\% & 4/19
& 21.1\% & 81.06 \\
Hard-pack (\texttt{n=20}) & Shenlong CGA & 19/20 & 95.0\% &
\textbf{9/20} & \textbf{45.0\%} & 51.89 \\
Hard-pack (\texttt{n=20}) & Simple CGA & \textbf{20/20} &
\textbf{100.0\%} & \textbf{9/20} & \textbf{45.0\%} & \textbf{39.20} \\
Hard-pack (\texttt{n=20}) & Euclidean 4×4 & \textbf{20/20} &
\textbf{100.0\%} & 5/20 & 25.0\% & 64.35 \\
\bottomrule
\end{tabular}
\end{table}

\begin{figure}
\centering
\includegraphics[width=\linewidth,keepaspectratio,trim=0bp 0bp 0bp 36bp,clip]{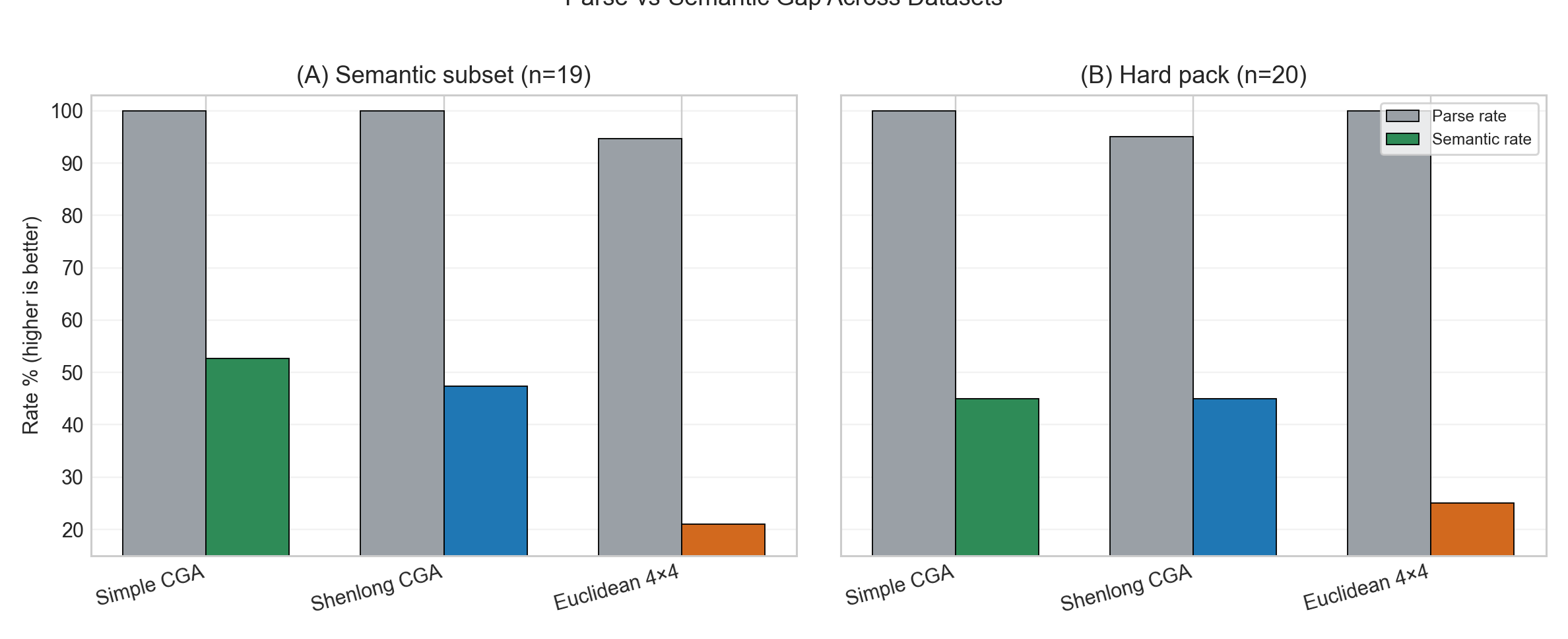}
\caption{Parse-semantic gap by dataset and method.}
\Description{Grouped comparison of parse-valid rates and semantic-valid rates across datasets and methods.}
\label{fig:10}
\end{figure}

Figure~\ref{fig:10} provides the corresponding visual summary of the
parse-semantic gap across the two evaluated datasets.

The parse-validity summary is therefore paired with explicit semantic
correctness metrics.

To test retry sensitivity, hard-pack runs are pooled by policy
(\texttt{k=1,2,3}) and compared in Table~\ref{tab:10}.

\begin{table}[htbp]
\centering
\caption{Hard-pack pass@k sensitivity (pooled repeats). Higher parse/semantic rates are better.}
\label{tab:10}
\begin{tabular}{@{}lllllll@{}}
\toprule
Approach & Parse@1 & Parse@2 & Parse@3 & Semantic@1 & Semantic@2 &
Semantic@3 \\
\midrule
Shenlong CGA & 95.0\% & 95.0\% & 95.0\% & \textbf{47.0\%} &
\textbf{45.0\%} & 44.0\% \\
Simple CGA & \textbf{100.0\%} & \textbf{100.0\%} & \textbf{100.0\%} &
45.0\% & \textbf{45.0\%} & \textbf{45.0\%} \\
Euclidean 4×4 & \textbf{100.0\%} & \textbf{100.0\%} & \textbf{100.0\%} &
25.0\% & 25.0\% & 25.0\% \\
\bottomrule
\end{tabular}
\end{table}

\begin{figure}
\centering
\pandocbounded{\includegraphics[keepaspectratio,trim=0bp 0bp 0bp 36bp,clip]{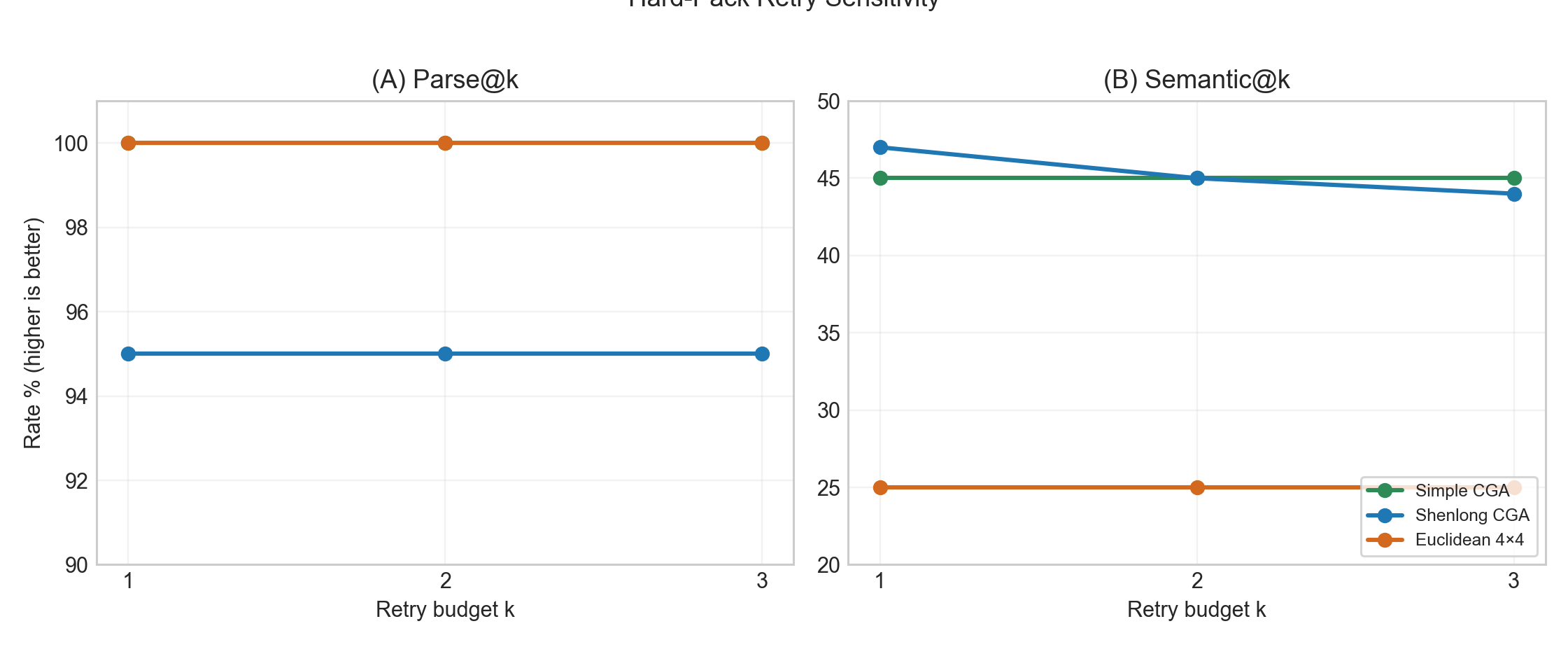}}
\caption{Hard-pack pass@k sensitivity curves.}
\Description{Curves showing how parse and semantic outcomes change under pass-at-k retry policies.}
\label{fig:11}
\end{figure}

Figure~\ref{fig:11} suggests that improving semantic grounding, rather
than format consistency alone, is the more relevant improvement
direction.

Pairwise Fisher exact tests on pooled repeats are summarized in
Table~\ref{tab:11}.

\begin{table}[htbp]
\centering
\caption{Hard-pack pairwise statistics (two-sided Fisher exact). Positive risk difference means Method A is higher; negative means Method A is lower. Significant p-values (\textless0.05) are shown in bold.}
\label{tab:11}
\small
\begin{tabular}{@{}p{0.16\linewidth}ccccp{0.30\linewidth}@{}}
\toprule
Comparison & \makecell{Parse Risk\\Diff (pp)} & \makecell{Parse\\p-value}
& \makecell{Semantic Risk\\Diff (pp)} & \makecell{Semantic\\p-value} &
Interpretation \\
\midrule
Euclidean vs Shenlong & +5 & 0.0594 & -22 to -19 & \textbf{0.0019 to
0.0072} & \makecell[l]{Euclidean shows lower semantic rates\\despite high parse rates.} \\
Euclidean vs Simple & +0 & 1.0000 & -20 & \textbf{0.0047} & \makecell[l]{Euclidean shows lower semantic rates\\than Simple CGA.} \\
Shenlong vs Simple & -5 & 0.0594 & -1 to +2 & 0.8872 to 1.0000 &
\makecell[l]{Hard-pack semantic rates are closely\\matched across CGA variants.} \\
\bottomrule
\end{tabular}
\end{table}

For completeness, Table~\ref{tab:12} reports the compact
SE3 control run (\texttt{n=20}) and the powered hard semantic suite
(\texttt{n=100} per method). Compact SE3 achieves complete parse
coverage and semantic rates in the same range as the CGA variants; on
token cost, it is the most compact representation in these runs.

\begin{table}[htbp]
\centering
\caption{Semantic follow-up datasets: compact non-CGA control and powered hard-suite results. The \texttt{n=100} block supports confirmatory inference for compact-vs-Euclidean contrasts (see Table~\ref{tab:13}).}
\label{tab:12}
\small
\begingroup
\setlength{\tabcolsep}{3pt}
\begin{tabular}{@{}p{0.18\linewidth}p{0.10\linewidth}p{0.09\linewidth}p{0.16\linewidth}p{0.09\linewidth}p{0.16\linewidth}p{0.12\linewidth}@{}}
\toprule
Dataset & Method & \makecell{Parse\\Success} & \makecell{Parse Rate\\(Wilson 95\% CI)} &
\makecell{Semantic\\Success} & \makecell{Semantic Rate\\(Wilson 95\% CI)} &
\makecell{Avg Tokens\\(parse-success\\rows)} \\
\midrule
Hard-pack + Compact SE3 (\texttt{n=20}) & Compact SE3 & \textbf{20/20} &
\makecell[l]{\textbf{100.0\%}\\\textbf{{[}83.9\%, 100.0\%{]}}} & \textbf{9/20} & \textbf{45.0\%
{[}25.8\%, 65.8\%{]}} & \textbf{23.65} \\
Hard-pack + Compact SE3 (\texttt{n=20}) & Euclidean 4×4 & 20/20 &
\makecell[l]{100.0\%\\{[}83.9\%, 100.0\%{]}} & 5/20 & 25.0\% {[}11.2\%, 46.9\%{]} &
64.15 \\
Hard-pack + Compact SE3 (\texttt{n=20}) & Shenlong CGA & 19/20 & \makecell[l]{95.0\%\\{[}76.4\%, 99.1\%{]}} & 9/20 & 45.0\% {[}25.8\%, 65.8\%{]} & 56.47 \\
Hard-pack + Compact SE3 (\texttt{n=20}) & Simple CGA & 20/20 & \makecell[l]{100.0\%\\{[}83.9\%, 100.0\%{]}} & 9/20 & 45.0\% {[}25.8\%, 65.8\%{]} & 36.85 \\
Powered hard suite (\texttt{n=100}) & Compact SE3 & \textbf{100/100} &
\makecell[l]{\textbf{100.0\%}\\\textbf{{[}96.3\%, 100.0\%{]}}} & 42/100 & 42.0\% {[}32.8\%,
51.8\%{]} & \textbf{21.45} \\
Powered hard suite (\texttt{n=100}) & Euclidean 4×4 & \textbf{100/100} &
\makecell[l]{\textbf{100.0\%}\\\textbf{{[}96.3\%, 100.0\%{]}}} & 24/100 & 24.0\% {[}16.7\%,
33.2\%{]} & 63.38 \\
Powered hard suite (\texttt{n=100}) & Shenlong CGA & 95/100 & \makecell[l]{95.0\%\\{[}88.8\%, 97.8\%{]}} & 44/100 & 44.0\% {[}34.7\%, 53.8\%{]} & 48.13 \\
Powered hard suite (\texttt{n=100}) & \textbf{Simple CGA} &
\textbf{100/100} & \makecell[l]{\textbf{100.0\%}\\\textbf{{[}96.3\%, 100.0\%{]}}} &
\textbf{45/100} & \textbf{45.0\% {[}35.6\%, 54.8\%{]}} & 37.34 \\
\bottomrule
\end{tabular}
\endgroup
\end{table}

In the powered suite (\texttt{n=100}), semantic pairwise contrasts
versus Euclidean are confirmatory: Simple CGA vs Euclidean is +21 pp
(\texttt{p=0.0028}), Shenlong vs Euclidean is +20 pp
(\texttt{p=0.0044}), and Compact SE3 vs Euclidean is +18 pp
(\texttt{p=0.0103}). Simple CGA and Compact SE3 are statistically close
on the same endpoint (+3 pp, \texttt{p=0.7755}), while Compact SE3 is
the lowest-token method.

To make inference strength explicit, Table~\ref{tab:13} reports semantic effect
sizes with 95\% CIs and p-values for the predefined contrasts.

\begin{table}[htbp]
\centering
\caption{Semantic effect sizes for predefined hard-suite contrasts. Positive risk difference favors the first method. Risk-difference CIs use normal approximation; relative-risk CIs use log-scale approximation; p-values are two-sided Fisher exact.}
\label{tab:13}
\small
\begin{tabular}{@{}p{0.14\linewidth}p{0.17\linewidth}p{0.14\linewidth}p{0.18\linewidth}p{0.14\linewidth}p{0.11\linewidth}@{}}
\toprule
Dataset & Comparison & \makecell{Semantic\\Success} & \makecell{Risk Difference\\(pp, 95\% CI)}
& \makecell{Relative Risk\\(95\% CI)} & \makecell{Fisher \texttt{p}\\(2-sided)} \\
\midrule
Hard-pack (\texttt{n=20}) & Compact SE3 vs Euclidean & 9/20 vs 5/20 &
+20.0 {[}-8.9, +48.9{]} & 1.80 {[}0.73, 4.43{]} & 0.3203 \\
Hard-pack (\texttt{n=20}) & Simple CGA vs Euclidean & 9/20 vs 5/20 &
+20.0 {[}-8.9, +48.9{]} & 1.80 {[}0.73, 4.43{]} & 0.3203 \\
Hard-pack (\texttt{n=20}) & Shenlong vs Euclidean & 9/20 vs 5/20 & +20.0
{[}-8.9, +48.9{]} & 1.80 {[}0.73, 4.43{]} & 0.3203 \\
Powered hard (\texttt{n=100}) & Compact SE3 vs Euclidean & 42/100 vs
24/100 & +18.0 {[}+5.2, +30.8{]} & 1.75 {[}1.15, 2.66{]} & 0.0103 \\
Powered hard (\texttt{n=100}) & Simple CGA vs Euclidean & 45/100 vs
24/100 & +21.0 {[}+8.1, +33.9{]} & 1.88 {[}1.24, 2.83{]} & 0.0028 \\
Powered hard (\texttt{n=100}) & Shenlong vs Euclidean & 44/100 vs 24/100
& +20.0 {[}+7.2, +32.8{]} & 1.83 {[}1.21, 2.77{]} & 0.0044 \\
Powered hard (\texttt{n=100}) & Simple CGA vs Compact SE3 & 45/100 vs
42/100 & +3.0 {[}-10.7, +16.7{]} & 1.07 {[}0.78, 1.47{]} & 0.7755 \\
\bottomrule
\end{tabular}
\end{table}

Power-aware interpretation: in the powered suite, compact-vs-Euclidean
semantic contrasts are both practically relevant (+18 to +21 pp) and
statistically confirmatory (\texttt{p\textless{}0.05}), while the
compact-vs-compact contrast is statistically close.

Adversarial-NL robustness is deferred to future work and is specified in
Section~\ref{7-conclusion} as a dedicated next-step evaluation item.

\subsection{Multi-Model
Replication}\label{56-multi-model-replication}

Table~\ref{tab:14} shows the protocol-fixed multi-model check
(\texttt{gpt-4o-mini}, \texttt{gpt-4.1-mini}) on hard-pack tasks
(\texttt{n=20} per model-method). Rates are directionally consistent
across the two tested closed models: Compact SE3 and Simple CGA both
exceed Euclidean on semantic success, and Compact SE3 is the
lowest-token method in both models. External validity is improved within
the closed-model family; open-model replication is the next expansion
step and is explicitly deferred (see
Section~\ref{7-conclusion}).

\begin{table}[htbp]
\centering
\caption{Multi-model hard-pack replication (protocol-fixed, \texttt{n=20} per model-method).}
\label{tab:14}
\small
\begin{tabular}{@{}p{0.12\linewidth}p{0.12\linewidth}ccccp{0.16\linewidth}@{}}
\toprule
Model & Method & \makecell{Parse\\Success} & \makecell{Parse\\Rate} &
\makecell{Semantic\\Success} & \makecell{Semantic\\Rate} &
\makecell{Avg Tokens\\(parse-success rows)} \\
\midrule
\texttt{gpt-4o-mini} & Compact SE3 & 20/20 & 100.0\% & \textbf{9/20} &
\textbf{45.0\%} & \textbf{21.25} \\
\texttt{gpt-4o-mini} & Euclidean 4×4 & 20/20 & 100.0\% & 5/20 & 25.0\% &
62.95 \\
\texttt{gpt-4o-mini} & Simple CGA & 20/20 & 100.0\% & \textbf{9/20} &
\textbf{45.0\%} & 37.55 \\
\texttt{gpt-4.1-mini} & Compact SE3 & 20/20 & 100.0\% & \textbf{11/20} &
\textbf{55.0\%} & \textbf{20.35} \\
\texttt{gpt-4.1-mini} & Euclidean 4×4 & 20/20 & 100.0\% & 5/20 & 25.0\%
& 63.75 \\
\texttt{gpt-4.1-mini} & Simple CGA & 20/20 & 100.0\% & \textbf{11/20} &
\textbf{55.0\%} & 34.00 \\
\bottomrule
\end{tabular}
\end{table}

Cross-model summary:

\begin{enumerate}
\def\labelenumi{\arabic{enumi}.}
\tightlist
\item
  Holds in both tested closed models: Euclidean is at 25.0\% hard-pack
  semantic success; Simple CGA and Compact SE3 are higher (45.0\% on
  \texttt{gpt-4o-mini}, 55.0\% on \texttt{gpt-4.1-mini}).
\item
  Holds in both tested closed models: Compact SE3 is the lowest-token
  method.
\item
  Not established by current data: stable separation between Simple CGA
  and Compact SE3, and transfer to open-weight model families.
\end{enumerate}

Open-weight replication is deferred to future work and is specified in
Section~\ref{7-conclusion} as the next external-validity expansion step.

Table~\ref{tab:15} summarizes key hard-suite trade-off metrics and
pairwise significance against Euclidean.

\begin{table}[htbp]
\centering
\caption{Cross-method hard-suite trade-off summary. Higher parse/semantic rates are better; lower tokens are better. \texttt{p\_sem\ vs\ Euclidean} compares each method with Euclidean within the same dataset. The hard-pack \texttt{n=20} rows are repeated from Table~\ref{tab:9} for direct side-by-side comparison with the powered block.}
\label{tab:15}
\small
\begin{tabular}{@{}p{0.14\linewidth}p{0.12\linewidth}p{0.16\linewidth}p{0.16\linewidth}c p{0.14\linewidth}@{}}
\toprule
Dataset & Method & \makecell{Parse Rate\\(95\% CI)} &
\makecell{Semantic Rate\\(95\% CI)} & \makecell{Avg\\Tokens} &
\makecell{\texttt{p\_\{sem\}}\\vs Euclidean} \\
\midrule
Hard-pack (\texttt{n=20}) & Compact SE3 & \makecell[l]{\textbf{100.0\%}\\\textbf{{[}83.9\%, 100.0\%{]}}} & \textbf{45.0\% {[}25.8\%, 65.8\%{]}} & \textbf{23.65} &
0.3203 \\
Hard-pack (\texttt{n=20}) & Euclidean 4×4 & \makecell[l]{\textbf{100.0\%}\\\textbf{{[}83.9\%, 100.0\%{]}}} & 25.0\% {[}11.2\%, 46.9\%{]} & 64.15 & - \\
Hard-pack (\texttt{n=20}) & Shenlong CGA & \makecell[l]{95.0\%\\{[}76.4\%, 99.1\%{]}} &
\textbf{45.0\% {[}25.8\%, 65.8\%{]}} & 56.47 & 0.3203 \\
Hard-pack (\texttt{n=20}) & Simple CGA & \makecell[l]{\textbf{100.0\%}\\\textbf{{[}83.9\%, 100.0\%{]}}} & \textbf{45.0\% {[}25.8\%, 65.8\%{]}} & 36.85 & 0.3203 \\
Powered hard (\texttt{n=100}) & Compact SE3 & \makecell[l]{\textbf{100.0\%}\\\textbf{{[}96.3\%, 100.0\%{]}}} & 42.0\% {[}32.8\%, 51.8\%{]} & \textbf{21.45} & 0.0103 \\
Powered hard (\texttt{n=100}) & Euclidean 4×4 & \makecell[l]{\textbf{100.0\%}\\\textbf{{[}96.3\%, 100.0\%{]}}} & 24.0\% {[}16.7\%, 33.2\%{]} & 63.38 & - \\
Powered hard (\texttt{n=100}) & Shenlong CGA & \makecell[l]{95.0\%\\{[}88.8\%, 97.8\%{]}} & 44.0\% {[}34.7\%, 53.8\%{]} & 48.13 & 0.0044 \\
Powered hard (\texttt{n=100}) & Simple CGA & \makecell[l]{\textbf{100.0\%}\\\textbf{{[}96.3\%, 100.0\%{]}}} & \textbf{45.0\% {[}35.6\%, 54.8\%{]}} & 37.34 & 0.0028 \\
\bottomrule
\end{tabular}
\end{table}

The main inference from Tables~\ref{tab:12}--\ref{tab:15} is that compact representations
(Simple CGA and Compact SE3) are both substantially cheaper than
Euclidean in token cost while remaining semantically stronger in the
powered hard suite (\texttt{n=100}). At the same time, the semantic gap
between the two compact variants is small (\texttt{p=0.7755} at
\texttt{n=100}), while sequence-fidelity still favors Simple CGA on the
dedicated compositional benchmark (Table~\ref{tab:6}).

\subsection{Interactive Latency and ENGAGE
Pilot}\label{57-interactive-latency-and-engage-pilot}

We ran a powered full-loop latency protocol on the hard instruction set
with \texttt{n=40} rows per method (20 tasks × 2 runs; 120 calls total)
under GPT-4o-mini and \texttt{retries=1}. Each call was timed end-to-end
from instruction dispatch to render-ready state and decomposed into
three components: LLM API time, parse/execute time, and render-ready
preparation time. Parse validity was 100\% for all three methods in this
protocol, so the latency decomposition is not confounded by parse
failures.

Table~\ref{tab:16} summarizes these results.

\begin{table}[htbp]
\centering
\caption{Full-loop latency decomposition for interactive editing (\texttt{n=40} per method). Lower values are better. Mean $\pm$ SD and median {[}IQR{]} are both reported because the Euclidean condition contains a long-tail API outlier. Headroom is relative to the 100ms target; update rate is the reciprocal of mean total latency.}
\label{tab:16}
\small
\begingroup
\setlength{\tabcolsep}{3pt}
\begin{tabular}{@{}p{0.09\linewidth}p{0.13\linewidth}p{0.13\linewidth}p{0.11\linewidth}p{0.11\linewidth}p{0.11\linewidth}p{0.10\linewidth}p{0.09\linewidth}@{}}
\toprule
Method & \makecell{Total Latency\\(s) mean $\pm$ SD} &
\makecell{Total Latency\\(s) median {[}IQR{]}} &
\makecell{API\\(s) mean $\pm$ SD} & \makecell{Parse+\\Execute\\(ms mean)} &
\makecell{Render-\\Ready\\(ms mean)} & \makecell{Headroom\\vs 100ms} &
\makecell{Max Update\\Rate (Hz)} \\
\midrule
\textbf{Compact SE3} & \textbf{0.94 $\pm$ 0.28} & \textbf{0.81 {[}0.38{]}} &
\textbf{0.94 $\pm$ 0.28} & 0.69 & 0.20 & \textbf{+0.84 s} & \textbf{1.06} \\
Simple CGA & 1.27 $\pm$ 0.38 & 1.19 {[}0.57{]} & 1.27 $\pm$ 0.38 & 1.04 & 0.42 &
+1.17 s & 0.79 \\
Euclidean 4×4 & 2.57 $\pm$ 4.91 & 1.40 {[}0.81{]} & 2.57 $\pm$ 4.90 &
\textbf{0.40} & \textbf{0.13} & +2.47 s & 0.39 \\
\bottomrule
\end{tabular}
\endgroup
\end{table}

The decomposition shows that latency is overwhelmingly API-bound for all
methods (\textgreater99.8\% of total mean time), while parse/execute and
render-ready overheads remain sub-millisecond. This indicates that
representation effects primarily act through token/API burden, not local
geometric execution cost. The Euclidean condition includes one long API
outlier (32.11 s), which inflates mean and SD; median total latencies
still preserve the same ordering (Compact SE3: 0.81 s, Simple CGA: 1.19
s, Euclidean 4×4: 1.40 s).

The minimal within-subject user pilot (ENGAGE-focused) is deferred to
immediate future work. The planned protocol is n=5--10 participants,
CGA-powered vs matrix-powered editing, primary endpoints task time and
error rate, secondary endpoint usability preference/SUS.

\subsection{Ablation Diagnostics}\label{58-ablation-diagnostics}

To isolate confounds, we run repeated ablations on a fixed 10-task set
under two policies (\texttt{pass@1}, \texttt{pass@2}) and four
conditions (Simple compact, Simple verbose, Shenlong, Euclidean),
yielding 50 evaluated rows per condition-policy. Full tables are
preserved in Appendix A (Tables~\ref{tab:a7}--\ref{tab:a8}); this section summarizes the key
diagnostic findings.

Table~\ref{tab:17} summarizes these results.

\begin{table}[htbp]
\centering
\caption{Repeated ablation summary (parse-focused; summary table). Parse rate is shown with Wilson CI (higher is better). Completion tokens and latency are cost metrics (lower is better). Shenlong\textquotesingle s lower parse rate relative to the Simple variants is partly associated with max-token budget effects under standard settings. Retry policy modestly shifts cost and Shenlong parse rates, while compact/verbose Simple variants remain parse-stable.}
\label{tab:17}
\begin{tabular}{@{}lllll@{}}
\toprule
Policy & Condition & Parse Rate (Wilson CI) & Avg Completion Tokens &
Avg Latency (s) \\
\midrule
pass@1 & Simple CGA (compact) & \textbf{100.0\% {[}92.9\%, 100.0\%{]}} &
36.72 & \textbf{1.29} \\
pass@1 & Simple CGA (verbose) & \textbf{100.0\% {[}92.9\%, 100.0\%{]}} &
\textbf{34.40} & 1.32 \\
pass@1 & Shenlong CGA & 80.0\% {[}67.0\%, 88.8\%{]} & 46.70 & 1.53 \\
pass@1 & Euclidean 4×4 & \textbf{100.0\% {[}92.9\%, 100.0\%{]}} & 71.58
& 1.89 \\
pass@2 & Simple CGA (compact) & \textbf{100.0\% {[}92.9\%, 100.0\%{]}} &
41.02 & \textbf{1.44} \\
pass@2 & Simple CGA (verbose) & \textbf{100.0\% {[}92.9\%, 100.0\%{]}} &
\textbf{33.54} & 1.45 \\
pass@2 & Shenlong CGA & 82.0\% {[}69.2\%, 90.2\%{]} & 46.34 & 1.68 \\
pass@2 & Euclidean 4×4 & \textbf{100.0\% {[}92.9\%, 100.0\%{]}} & 74.50
& 1.79 \\
\bottomrule
\end{tabular}
\end{table}

The ablation evidence supports two conservative claims: (i)
Shenlong\textquotesingle s parse-rate headroom stays under modest retry
increases, and this deficit should be attributed to its prompt-length
budget interaction rather than to any fundamental CGA representational
weakness; (ii) representation-level and prompt-level effects interact,
so token count alone should not be treated as a single explanatory
variable. Full pairwise statistics are in Appendix A (Table~\ref{tab:a8}).

\begin{figure}
\centering
\pandocbounded{\includegraphics[keepaspectratio,trim=0bp 0bp 0bp 56bp,clip]{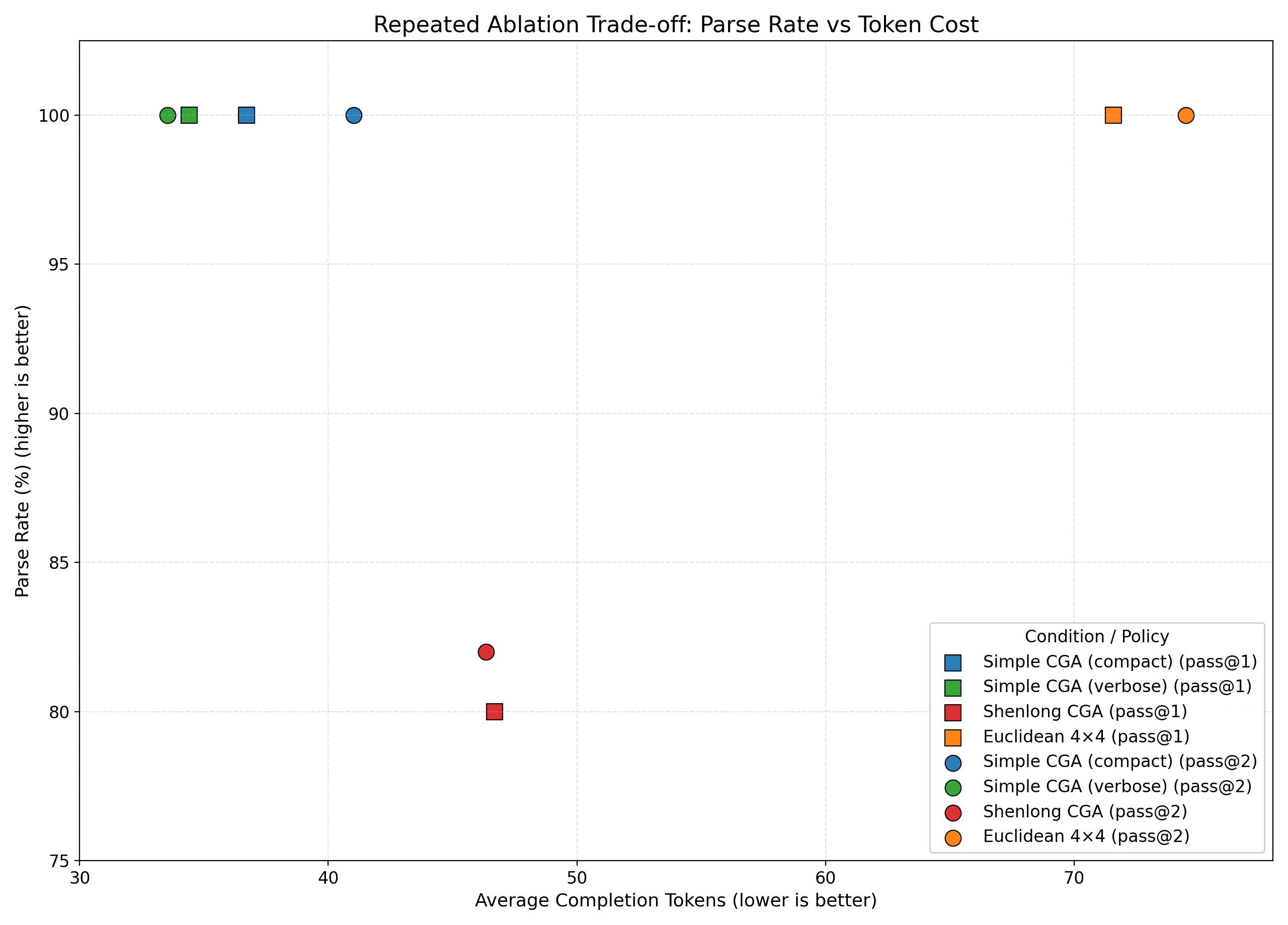}}
\caption{Ablation parse-rate vs completion-token trade-off.}
\Description{Ablation trade-off plot showing parse-rate performance versus completion-token cost.}
\label{fig:12}
\end{figure}

Figure~\ref{fig:12} shows the Simple variants on a stronger
parse-rate/cost frontier, while Euclidean shifts rightward in completion
cost.

Interpretation limits: several subsets are modest in size (semantic
subset \texttt{n=19}, hard-pack \texttt{n=20}), and policy choice
(\texttt{pass@1} vs \texttt{pass@k}) affects parse-oriented outcomes
(Tables~\ref{tab:10} and~\ref{tab:17}). Accordingly, rates are reported as
protocol-conditional evidence under the tested model family, prompts,
and retry settings, rather than universal performance constants.

\subsection{Reproducibility and Supplementary
Material}\label{59-reproducibility-and-supplementary-material}

To keep the main narrative readable while preserving traceability, the
following supporting assets are moved to Appendix A:

\begin{itemize}
\tightlist
\item
  protocol snapshot (Table~\ref{tab:a1});
\item
  detailed per-task core tables and supplementary figures
  (Tables~\ref{tab:a2}--\ref{tab:a5},
  Figures~\ref{fig:a1}--\ref{fig:a3});
\item
  compact visual summaries are shown in main text
  (Figures~\ref{fig:7}--\ref{fig:12}), while
  full supporting numeric rows remain in Appendix A tables;
\item
  follow-up datasets (hard-pack + Compact SE3, multi-model hard-pack,
  expanded hard-suite uncertainty) are included in the supplementary
  materials;
\item
  hard-pack retry-aware cost table (Table~\ref{tab:a6});
\item
  single-run ablation and full repeated-ablation pairwise matrix
  (Tables~\ref{tab:a7}--\ref{tab:a8});
\item
  consolidated supplementary completeness tables
  (Tables~\ref{tab:a9}--\ref{tab:a10}).
\end{itemize}

\begin{center}\rule{0.5\linewidth}{0.5pt}\end{center}

\section{Discussion}\label{6-discussion}

\subsection{Sequence-Fidelity Evidence for Algebraic
Expressiveness}\label{61-sequence-fidelity-as-mechanistic-evidence-for-algebraic-expressiveness}

The sequence-fidelity result (Table~\ref{tab:6},
Section~\ref{52-sequence-fidelity-stress-result-primary-confirmatory-finding})
shows a +7.5 pp advantage for Simple CGA over Compact SE3
(\texttt{p=0.016}) under parse saturation for both methods, with a modest
token gap (21 tokens). Because both methods are compact and both parse at
100\%, this contrast isolates ordered-chain behavior rather than
parseability.

A plausible account is that CGA motor composition encodes a chain as a
single algebraic object, whereas Compact SE3 uses list-position semantics
across typed JSON operations. The data are consistent with this
hypothesis, but they do not by themselves establish a causal mechanism.
Latency should be read as a separate axis in Table~\ref{tab:16}: Compact
SE3 is faster on mean latency, while Simple CGA retains the
sequence-fidelity advantage.

\subsection{Semantic Trends for Compact
Interfaces}\label{62-semantic-trends-as-directional-evidence-for-compact-over-verbose}

Hard-pack and semantic checks show that parse success is only a partial
proxy for geometric task success (Tables~\ref{tab:9}
and~\ref{tab:10}), so syntax-valid outputs and geometry-valid outcomes
should be reported separately.

Under that criterion, the powered hard-suite results (\texttt{n=100})
show that all three compact interfaces (Simple CGA, Shenlong CGA, and
the Compact SE3 control) outperform the Euclidean baseline on semantic
success (Table~\ref{tab:13}, \texttt{p\textless{}0.05} in all three
compact-vs-Euclidean contrasts). The Simple CGA vs Compact SE3 semantic
contrast remains statistically close (\texttt{p=0.7755}), so the
supported aggregate claim is compact-interface advantage over the tested
Euclidean baseline, with a Simple-CGA advantage confined to the
sequence-fidelity endpoint.

\subsection{Latency and Interactive
Viability}\label{63-latency-and-interactive-viability}

The token-efficiency story, when translated to wall-clock latency
(Table~\ref{tab:3}), has direct relevance for immersive NL-editing
interfaces. The
full-loop measurements in
Section~\ref{57-interactive-latency-and-engage-pilot} show Compact SE3
as fastest (0.94 $\pm$ 0.28
s), followed by Simple CGA (1.27 $\pm$ 0.38 s), with Euclidean 4×4 slower on
mean latency (2.57 $\pm$ 4.91 s). Relative to a 100ms target, the measured
headroom is +0.84 s, +1.17 s, and +2.47 s, respectively.

All tested methods are currently above the 100ms sub-interactive
threshold. The gap is dominated by LLM API latency rather than
representation overhead; architectural work on streaming execution,
caching, or local inference would need to close an order of magnitude.
In this protocol, API time contributes more than 99.8\% of total latency
for every method, while parse/execute and render-ready stages are
sub-millisecond. The representation choice still matters because it
modulates token/API burden: both compact interfaces (Compact SE3, Simple
CGA) retain clear latency advantages over the verbose Euclidean
baseline, and these differences accumulate over multi-step edits.

\subsection{Scope, Validity, and Practical
Implications}\label{64-scope-validity-and-practical-implications}

The claims are valid for the evaluated system: GPT-4o-mini /
GPT-4.1-mini, the current template-execution pipeline, primitive-shape
scenes, and the tested task distributions. Immediate next-step
validation includes transfer tests to additional model families,
open-world 3D assets, and more unconstrained instructions. The matched
closed-model check (Table~\ref{tab:14}) supports directional consistency within
the GPT family and provides a concrete launch point for broader
cross-provider transfer studies.

Within those bounds, the practical implication is clear: choosing a
compact, closed-form geometric interface is a high-leverage design
decision for LLM-to-geometry systems. CGA is particularly attractive
when unified algebraic composition is required --- specifically, when
ordered instruction chains and motor composition semantics are desirable
--- and the current evidence motivates broader comparative studies
across compact symbolic encodings.

\subsection{Threats to Validity}\label{65-threats-to-validity}

Three named threats are identified proactively:

\textbf{Threat 1: Training-data contamination.} GPT-family models may
have encountered CGA syntax during pre-training (e.g., from academic
code repositories or tutorial pages for the \texttt{clifford} or
\texttt{kingdon} Python libraries). If the model has memorized CGA
expression patterns, its "generation" of CGA motors may partially
reflect recall rather than compositional reasoning, inflating
CGA\textquotesingle s apparent advantage over less well-represented
formalisms. Mitigation: open-weight model replication
(Section~\ref{7-conclusion}) on models with more transparent
training corpora would
partially address this threat. The Compact SE3 control baseline, being a
bespoke JSON schema unlikely to appear in pre-training data, provides
some protection against this confound in the semantic suite; the
sequence-fidelity result
(Section~\ref{52-sequence-fidelity-stress-result-primary-confirmatory-finding})
uses the same Compact SE3 control for a
direct compact-vs-compact comparison that is less susceptible to
recall-based inflation.

\textbf{Threat 2: Compact SE3 underspecification risk.} Comparative
studies can under-specify compact non-CGA baselines, making fairness
difficult to assess. This paper mitigates that risk by providing a full
Compact SE3 schema and execution semantics in
Section~\ref{44-compact-se3-baseline--full-specification}. A remaining
limitation is that per-task schema interpretation traces are not
reported in the main text.

\textbf{Threat 3: Shenlong max-tokens budget artifact.} The Shenlong CGA
variant\textquotesingle s lower parse rates (80--82\% in ablation, Table
17) are partly attributable to a max-tokens budget interaction: its
verbose chain-of-thought prompt generates 200--500 additional reasoning
tokens that can push total output past the token limit, causing
truncated JSON. This makes Shenlong\textquotesingle s apparent
disadvantage relative to Simple CGA a partially unfair comparison at
standard token budgets. Reporting Shenlong as a legacy engineering
reference point rather than a primary competitor (as done in
Section~\ref{43-three-competing-prompting-strategies}) is
the editorial response; a fairness run with a larger output budget would
allow a cleaner statement.

\begin{center}\rule{0.5\linewidth}{0.5pt}\end{center}

\section{Conclusion}\label{7-conclusion}

This paper evaluates whether output representation changes the
reliability and compositional fidelity of LLM-driven 3D scene editing.
Our strongest result is confirmatory: under a sequence-stress protocol
designed to isolate ordered-chain faithfulness (\texttt{n=120} outputs
per method), Simple CGA preserves exact operation sequences more
reliably than a Compact SE3 control (97.5\% vs 90.0\%,
\texttt{p=0.016}) while using fewer completion tokens (112.6 vs 133.6).
This result holds under parse saturation (both methods at 100\% parse
rate), and is consistent with the hypothesis that algebraic expression
form --- not compactness alone --- is associated with higher
compositional faithfulness.

A second result is now confirmatory on the powered semantic suite
(\texttt{n=100} per method): compact representations (Simple CGA 45.0\%,
Compact SE3 42.0\%, Shenlong 44.0\%) exceed the Euclidean 4×4 baseline
(24.0\%), with effect sizes of +21 pp (\texttt{p=0.0028}) for Simple CGA
vs Euclidean, +18 pp (\texttt{p=0.0103}) for Compact SE3 vs Euclidean,
and +20 pp (\texttt{p=0.0044}) for Shenlong vs Euclidean. Observed
effect sizes are practically relevant, while the Simple CGA vs Compact
SE3 semantic difference is statistically close (+3 pp,
\texttt{p=0.7755}). The protocol is fixed and the compact-vs-Euclidean
direction is consistent across two closed model families.

A third contribution is methodological. The consistent gap between parse
success and semantic success across all tested methods and datasets
indicates that syntax-valid outputs are a necessary but insufficient
proxy for geometric correctness. Future evaluation protocols for
LLM-to-geometry systems should explicitly separate these two criteria.

Immediate future work should prioritise four directions. First, run an
adversarial natural-language robustness subset (10-20 instructions
authored without representation hints) under the fixed protocol and
report method-wise semantic rates as a dedicated robustness endpoint.
Second, apply grammar-constrained CGA decoding --- using structured
generation frameworks such as Outlines or Guidance to enforce the CGA
expression grammar by construction --- to eliminate residual parse
failures and isolate semantic grounding as the remaining frontier.
Third, perform open-weight model replication (e.g., Llama-3/Mistral
families) under matched settings to test whether the compact-vs-verbose
trend transfers beyond the GPT model family and to reduce
training-data-contamination concerns. Fourth, add closed-loop visual
verification --- coupling the execution engine to a rendering oracle
that scores geometric satisfaction automatically --- as a scalable path
to larger semantic suites for confirmatory inference. Together these
steps outline a path from the current evidence to more robust,
model-general conclusions about symbolic interface design for
LLM-to-geometry pipelines.

\section{Acknowledgments}\label{acknowledgments}

This work was partially supported by the Innosuisse OMEN-E project and
by the European Union\textquotesingle s research and innovation
programs. The authors thank the Kingdon library developers for the
open-source Geometric Algebra implementation~\cite{roelfs2025willingkingdoncliffordalgebra} that made this work
possible.

\begin{center}\rule{0.5\linewidth}{0.5pt}\end{center}

\section{Disclosure of Interests}\label{disclosure-of-interests}

The authors declare the following interests: G. Papagiannakis is the
founder and CEO of ORamaVR SA, which develops the MAGES SDK platform
referenced in this paper.

\begin{center}\rule{0.5\linewidth}{0.5pt}\end{center}

\bibliographystyle{ACM-Reference-Format}
\bibliography{references}

\begin{center}\rule{0.5\linewidth}{0.5pt}\end{center}

\section{Appendix A: Supplementary Results and Validation
Tables}\label{appendix-a-supplementary-results-and-audit-tables}

This appendix preserves detailed outputs that were removed from the main
Section~\ref{5-experiments-and-results} for readability. No experimental
results are discarded.

\subsection{Protocol Snapshot}\label{a1-protocol-snapshot}

\begin{table}[htbp]
\centering
\caption{Protocol snapshot and fixed settings used by the main benchmarks, including model selection, prompt lengths, retry policy, and token budgets for the core benchmark, powered semantic suite, and powered latency protocol.}
\label{tab:a1}
\small
\begin{tabular}{@{}p{0.34\linewidth}p{0.60\linewidth}@{}}
\toprule
Field & Value \\
\midrule
Snapshot timestamp & 2026-04-16 19:12:37 \\
Model & gpt-4o-mini \\
\makecell[l]{Prompt length\\(Simple/Shenlong/Euclidean/Compact SE3)} & 721 / 963 / 435
/ 588 characters \\
Temperature schedule & 0.1 + 0.05 * attempt \\
Retry policy (defaults) & \texttt{run\_one} retries = 2; stress max
shots = 3 \\
Max tokens per core-33 benchmark block & 5-object: 500/300/400, stress:
500/500/500, 10-object: 500/400/500, accuracy: 300/300/300, 100-object:
600/600/600 (Shenlong/Simple/Euclidean) \\
Max tokens (powered semantic suite,
Section~\ref{55-semantic-validity-under-harder-grounding}) &
model: gpt-4o-mini;
\texttt{n=100} per method (20 tasks × 5 variants); retries=1;
max\_tokens: 600 (Simple/Shenlong/Euclidean), 500 (Compact SE3) \\
Powered latency protocol settings
(Section~\ref{57-interactive-latency-and-engage-pilot}) &
model: gpt-4o-mini;
\texttt{n=40} per method (20 tasks × 2 runs); retries=1; max\_tokens:
600 (Simple/Euclidean), 500 (Compact SE3) \\
\bottomrule
\end{tabular}
\end{table}

\subsection{Detailed Core Benchmark
Tables}\label{a2-detailed-core-benchmark-tables}

\begin{table}[htbp]
\centering
\caption{Task-level 5-object benchmark outcomes across methods, with per-task completion-token counts. \texttt{OK} denotes parse and semantic success; \texttt{FAIL} denotes parse failure or semantic failure.}
\label{tab:a2}
\begin{tabular}{@{}lllllll@{}}
\toprule
Task Type & Shenlong & Tokens & Simple CGA & Tokens & Euclidean &
Tokens \\
\midrule
Simple translate & OK & 127 & OK & 25 & OK & 41 \\
Stacking & OK & 33 & OK & 29 & OK & 45 \\
Scaling & OK & 15 & OK & 12 & FAIL & 0 \\
Rotation & FAIL & 0 & OK & 20 & OK & 41 \\
T+R+D compose & OK & 49 & OK & 47 & OK & 55 \\
Stack+scale & OK & 38 & OK & 34 & OK & 41 \\
Multi-object & OK & 45 & OK & 77 & OK & 123 \\
Hard spatial & FAIL & 0 & OK & 22 & OK & 57 \\
\textbf{Total} & \textbf{6/8 (75\%)} & \textbf{avg 51} & \textbf{8/8
(100\%)} & \textbf{avg 33} & \textbf{7/8 (88\%)} & \textbf{avg 58} \\
\bottomrule
\end{tabular}
\end{table}

\begin{table}[htbp]
\centering
\caption{Task-level stress-test outcomes under \texttt{pass@3}, showing per-task success/failure by method across compositional and robustness-oriented instruction types.}
\label{tab:a3}
\begin{tabular}{@{}llll@{}}
\toprule
Task Type & Shenlong & Simple CGA & Euclidean \\
\midrule
Irrational angle & OK & OK & OK \\
Triple compose & OK & OK & OK \\
5-obj edit & OK & OK & OK \\
Chained rotation & OK & OK & OK \\
Scale all & OK & OK & FAIL \\
Stack+relative & OK & OK & OK \\
\textbf{Total} & \textbf{6/6} & \textbf{6/6} & \textbf{5/6} \\
\bottomrule
\end{tabular}
\end{table}

\begin{table}[htbp]
\centering
\caption{Expanded spatial-accuracy suite (\texttt{n=18}) with inferential summary. Exact success uses threshold error \texttt{\textless{}0.5}; Wilson 95\% CIs are shown as {[}lower, upper{]}. Lower errors are better.}
\label{tab:a4}
\small
\begin{tabular}{@{}p{0.14\linewidth}p{0.11\linewidth}p{0.20\linewidth}p{0.10\linewidth}p{0.10\linewidth}p{0.19\linewidth}@{}}
\toprule
Method & \makecell{Exact\\Success} & \makecell{Exact Rate\\(Wilson 95\% CI)} &
\makecell{Mean\\Error} & \makecell{Median\\Error} &
\makecell{Avg Tokens\\(parse-success rows)} \\
\midrule
Shenlong CGA & \textbf{18/18} & \textbf{100.0\% {[}82.4\%, 100.0\%{]}} &
\textbf{0.00} & \textbf{0.00} & 44.56 \\
\textbf{Simple CGA} & \textbf{18/18} & \textbf{100.0\% {[}82.4\%,
100.0\%{]}} & \textbf{0.00} & \textbf{0.00} & \textbf{29.50} \\
Euclidean 4×4 & 5/18 & 27.8\% {[}12.5\%, 50.9\%{]} & 2.35 & 3.30 &
47.11 \\
\bottomrule
\end{tabular}
\end{table}

Pairwise exact-success tests (two-sided Fisher) on this suite: Euclidean
vs Shenlong \texttt{p\textless{}1e-5}; Euclidean vs Simple CGA
\texttt{p\textless{}1e-5}; Shenlong vs Simple CGA \texttt{p=1.0}.

\begin{table}[htbp]
\centering
\caption{100-object benchmark summary under the 30-object context setting, reporting method-wise success and average completion-token cost.}
\label{tab:a5}
\begin{tabular}{@{}lll@{}}
\toprule
Approach & Success & Avg Tokens \\
\midrule
Shenlong CGA & 9/10 (90\%) & 64 \\
Simple CGA & 10/10 (100\%) & 47 \\
Euclidean 4×4 & 10/10 (100\%) & 86 \\
\bottomrule
\end{tabular}
\end{table}

Supplementary full-context run (all 100 object descriptions): Shenlong
9/10 (avg 64), Simple CGA 10/10 (avg 50), Euclidean 10/10 (avg 85).

\begin{figure}
\centering
\pandocbounded{\includegraphics[keepaspectratio]{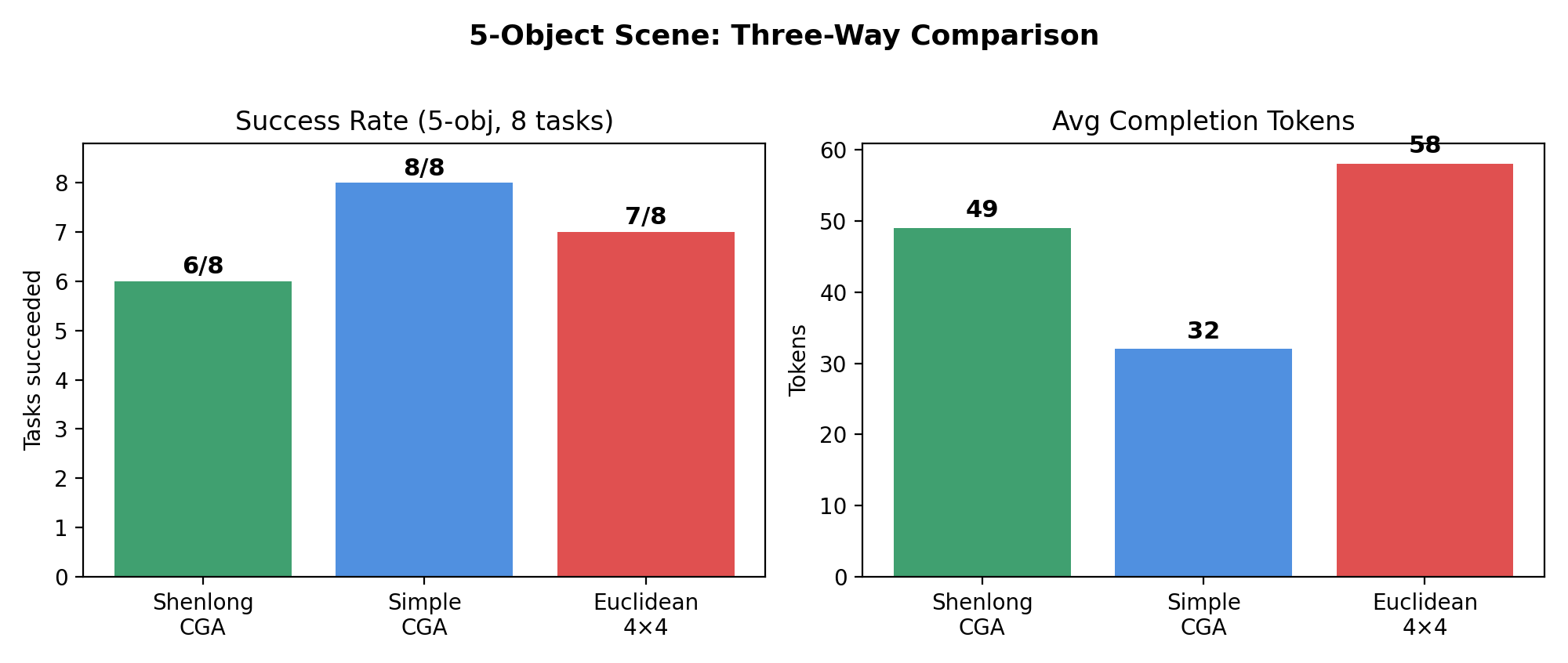}}
\caption{5-object scene comparison.}
\Description{Appendix chart comparing methods on the 5-object benchmark scenario.}
\label{fig:a1}
\end{figure}

Figure~\ref{fig:a1} compares 5-object outcomes; higher success and lower
tokens are preferable.

\begin{figure}
\centering
\pandocbounded{\includegraphics[keepaspectratio]{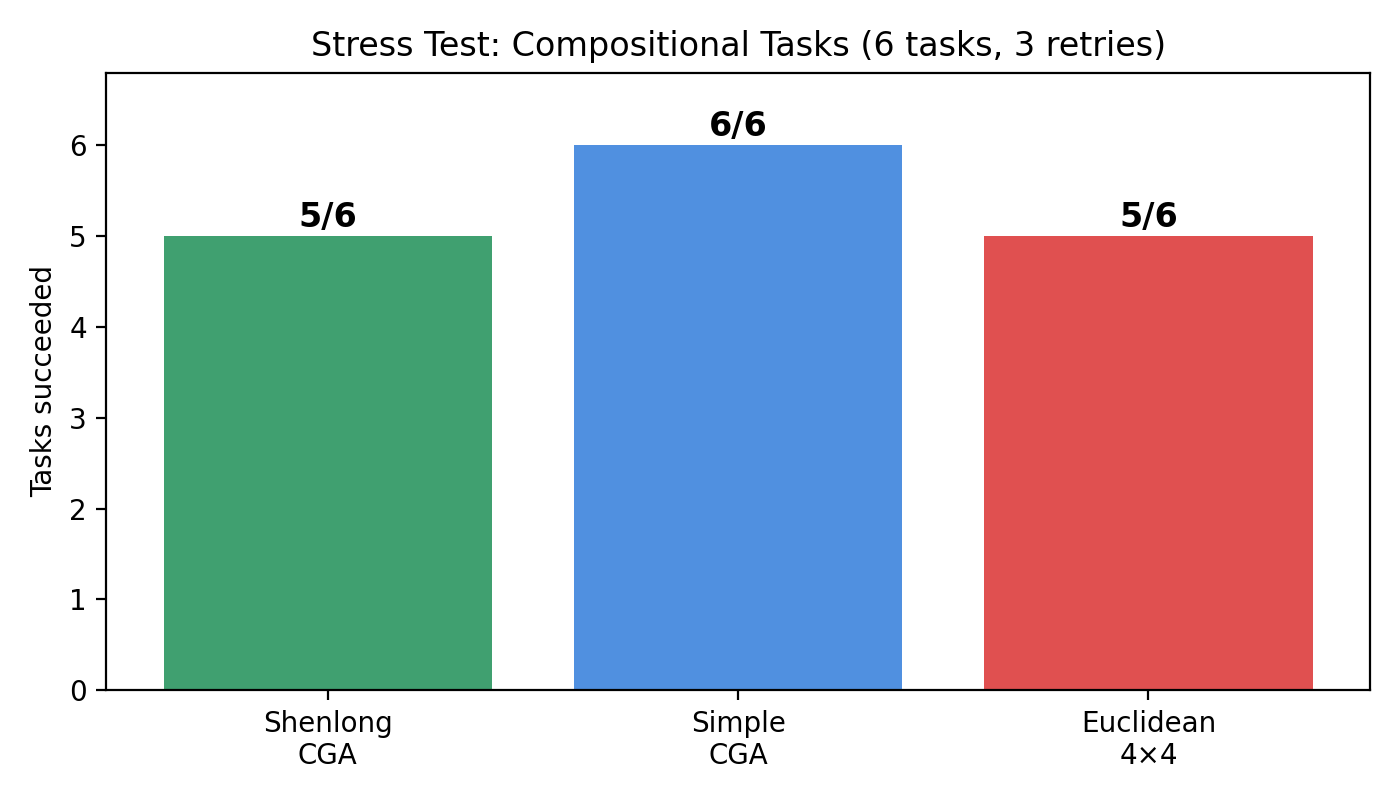}}
\caption{Stress-test comparison.}
\Description{Appendix chart comparing method performance on stress-test tasks.}
\label{fig:a2}
\end{figure}

Figure~\ref{fig:a2} summarizes stress-test behavior under the same
instruction structure.

\begin{figure}
\centering
\pandocbounded{\includegraphics[keepaspectratio]{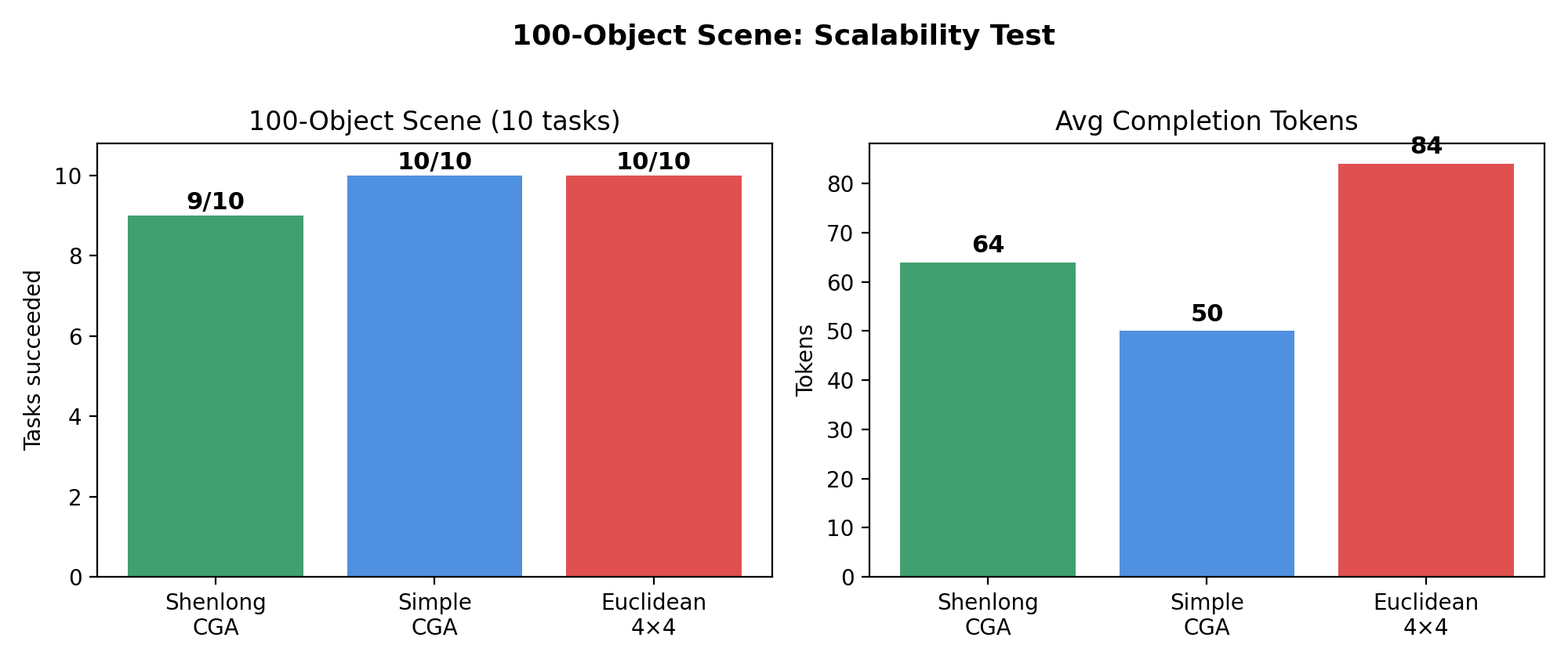}}
\caption{100-object benchmark comparison.}
\Description{Appendix chart comparing methods on the 100-object benchmark.}
\label{fig:a3}
\end{figure}

Figure~\ref{fig:a3} compares 100-object benchmark outcomes, where lower
token cost is better at matched success.

\subsection{Cost, Ablation, and Verification Support
Tables}\label{a3-cost-ablation-and-audit-support-tables}

\begin{table}[htbp]
\centering
\caption{Hard-pack retry-aware cost and latency by method and retry budget (\texttt{k}), reporting average completion tokens (success rows), average total tokens across all attempts, and average per-task latency.}
\label{tab:a6}
\small
\begin{tabular}{@{}c l p{0.23\linewidth}p{0.23\linewidth}p{0.13\linewidth}@{}}
\toprule
Retries (k) & Approach & \makecell{Avg Completion Tokens\\(success rows)} &
\makecell{Avg Total Tokens\\(all attempts)} &
\makecell{Avg Latency\\per Task (s)} \\
\midrule
1 & Shenlong CGA & 52.48 & 601.26 & 1.40 \\
1 & Simple CGA & 37.99 & 513.09 & 1.24 \\
1 & Euclidean 4×4 & 64.24 & 456.34 & 1.57 \\
2 & Shenlong CGA & 54.34 & 633.32 & 2.29 \\
2 & Simple CGA & 38.35 & 513.45 & 1.27 \\
2 & Euclidean 4×4 & 63.77 & 455.87 & 1.56 \\
3 & Shenlong CGA & 50.65 & 660.12 & 1.55 \\
3 & Simple CGA & 38.00 & 513.10 & 1.30 \\
3 & Euclidean 4×4 & 64.62 & 456.72 & 1.58 \\
\bottomrule
\end{tabular}
\end{table}

\begin{table}[htbp]
\centering
\caption{Single-run ablation summary by policy and condition, reporting tasks attempted, successes, success rates, and average completion tokens for successful rows.}
\label{tab:a7}
\begin{tabular}{@{}llllll@{}}
\toprule
Policy & Condition & Tasks & Success & Success Rate & Avg Tokens
(success rows) \\
\midrule
pass@1 & Simple CGA (compact) & 10 & 10 & 100.0\% & 42.40 \\
pass@1 & Simple CGA (verbose) & 10 & 10 & 100.0\% & 35.50 \\
pass@1 & Shenlong CGA & 10 & 8 & 80.0\% & 49.25 \\
pass@1 & Euclidean 4×4 & 10 & 10 & 100.0\% & 74.00 \\
pass@2 & Simple CGA (compact) & 10 & 10 & 100.0\% & 60.30 \\
pass@2 & Simple CGA (verbose) & 10 & 10 & 100.0\% & 35.40 \\
pass@2 & Shenlong CGA & 10 & 8 & 80.0\% & 51.25 \\
pass@2 & Euclidean 4×4 & 10 & 10 & 100.0\% & 71.10 \\
\bottomrule
\end{tabular}
\end{table}

\begin{table}[htbp]
\centering
\caption{Full repeated-ablation pairwise matrix across \texttt{pass@1} and \texttt{pass@2}, including rate contrasts, risk difference, relative risk, odds ratio, and two-sided Fisher p-values.}
\label{tab:a8}
\small
\begingroup
\setlength{\tabcolsep}{2pt}
\begin{tabular}{@{}p{0.06\linewidth}>{\centering\arraybackslash}p{0.06\linewidth}p{0.12\linewidth}p{0.12\linewidth}p{0.06\linewidth}p{0.06\linewidth}>{\centering\arraybackslash}p{0.10\linewidth}>{\centering\arraybackslash}p{0.10\linewidth}p{0.08\linewidth}>{\centering\arraybackslash}p{0.12\linewidth}@{}}
\toprule
Policy & \makecell{Retries\\(k)} & Condition A & Condition B &
\makecell{Rate\\A} & \makecell{Rate\\B} &
\makecell{Risk\\Difference} & \makecell{Relative\\Risk} &
\makecell{Odds\\Ratio} & \makecell{Fisher\\p-value\\(2-sided)} \\
\midrule
pass@1 & 1 & Simple CGA (compact) & Simple CGA (verbose) & 100.0\% &
100.0\% & +0.0 pp & 1.000 & 1.000 & 1.0000 \\
pass@1 & 1 & Simple CGA (compact) & Shenlong CGA & 100.0\% & 80.0\% &
+20.0 pp & 1.247 & 26.185 & 0.0012 \\
pass@1 & 1 & Simple CGA (compact) & Euclidean 4×4 & 100.0\% & 100.0\% &
+0.0 pp & 1.000 & 1.000 & 1.0000 \\
pass@1 & 1 & Simple CGA (verbose) & Shenlong CGA & 100.0\% & 80.0\% &
+20.0 pp & 1.247 & 26.185 & 0.0012 \\
pass@1 & 1 & Simple CGA (verbose) & Euclidean 4×4 & 100.0\% & 100.0\% &
+0.0 pp & 1.000 & 1.000 & 1.0000 \\
pass@1 & 1 & Shenlong CGA & Euclidean 4×4 & 80.0\% & 100.0\% & -20.0 pp
& 0.802 & 0.038 & 0.0012 \\
pass@2 & 2 & Simple CGA (compact) & Simple CGA (verbose) & 100.0\% &
100.0\% & +0.0 pp & 1.000 & 1.000 & 1.0000 \\
pass@2 & 2 & Simple CGA (compact) & Shenlong CGA & 100.0\% & 82.0\% &
+18.0 pp & 1.217 & 23.120 & 0.0026 \\
pass@2 & 2 & Simple CGA (compact) & Euclidean 4×4 & 100.0\% & 100.0\% &
+0.0 pp & 1.000 & 1.000 & 1.0000 \\
pass@2 & 2 & Simple CGA (verbose) & Shenlong CGA & 100.0\% & 82.0\% &
+18.0 pp & 1.217 & 23.120 & 0.0026 \\
pass@2 & 2 & Simple CGA (verbose) & Euclidean 4×4 & 100.0\% & 100.0\% &
+0.0 pp & 1.000 & 1.000 & 1.0000 \\
pass@2 & 2 & Shenlong CGA & Euclidean 4×4 & 82.0\% & 100.0\% & -18.0 pp
& 0.822 & 0.043 & 0.0026 \\
\bottomrule
\end{tabular}
\endgroup
\end{table}

\begin{table}[htbp]
\centering
\caption{Consolidated supplementary completeness checklist confirming the presence of protocol, uncertainty, semantic, ablation, and pairwise-reporting artifacts in the supplementary material.}
\label{tab:a9}
\small
\begin{tabular}{@{}p{0.65\linewidth}p{0.25\linewidth}@{}}
\toprule
Checklist Item & Status \\
\midrule
Protocol snapshot present & Yes \\
Core pass-policy summary present & Yes \\
Core uncertainty summary present & Yes \\
Semantic subset summary present & Yes \\
Ablation summary present & Yes \\
Hard-pack summary present & Yes \\
Hard-pack rows present & Yes \\
Pairwise rows written & 6 \\
\bottomrule
\end{tabular}
\end{table}

\begin{table}[htbp]
\centering
\caption{Single-run hard-pack pairwise cross-check (\texttt{n=20} per method), listing parse and semantic pairwise contrasts with rate differences, relative risk, odds ratio, and two-sided Fisher p-values.}
\label{tab:a10}
\small
\begingroup
\setlength{\tabcolsep}{2pt}
\begin{tabular}{@{}p{0.10\linewidth}p{0.07\linewidth}p{0.10\linewidth}p{0.10\linewidth}p{0.07\linewidth}p{0.07\linewidth}p{0.09\linewidth}p{0.09\linewidth}p{0.08\linewidth}p{0.09\linewidth}@{}}
\toprule
Dataset & Metric & Method A & Method B & \makecell{Rate\\A} &
\makecell{Rate\\B} & \makecell{Risk\\Difference} &
\makecell{Relative\\Risk} & \makecell{Odds\\Ratio} &
\makecell{Fisher\\p-value\\(2-sided)} \\
\midrule
\makecell[l]{hardpack\\latest} & parse & Euclidean & Shenlong & 100.0\% & 95.0\% &
+5.0 pp & 1.051 & 3.154 & 1.0000 \\
\makecell[l]{hardpack\\latest} & parse & Euclidean & Simple CGA & 100.0\% & 100.0\% &
+0.0 pp & 1.000 & 1.000 & 1.0000 \\
\makecell[l]{hardpack\\latest} & parse & Shenlong & Simple CGA & 95.0\% & 100.0\% &
-5.0 pp & 0.951 & 0.317 & 1.0000 \\
\makecell[l]{hardpack\\latest} & \makecell[l]{seman\\tic} & Euclidean & Shenlong & 25.0\% & 45.0\% &
-20.0 pp & 0.579 & 0.430 & 0.3203 \\
\makecell[l]{hardpack\\latest} & \makecell[l]{seman\\tic} & Euclidean & Simple CGA & 25.0\% & 45.0\% &
-20.0 pp & 0.579 & 0.430 & 0.3203 \\
\makecell[l]{hardpack\\latest} & \makecell[l]{seman\\tic} & Shenlong & Simple CGA & 45.0\% & 45.0\% &
+0.0 pp & 1.000 & 1.000 & 1.0000 \\
\bottomrule
\end{tabular}
\endgroup
\end{table}

\begin{center}\rule{0.5\linewidth}{0.5pt}\end{center}

\section{Appendix B: Implementation
Details}\label{appendix-b-implementation-details}

\subsection{CGA Primitives
Implementation}\label{b1-cga-primitives-implementation}

The CGA primitives are implemented using the Kingdon library for
Python~\cite{roelfs2025willingkingdoncliffordalgebra}.
The algebra is constructed as \texttt{Algebra(4,\ 1)}, yielding the
32-dimensional Cl(4,1) space. Key implementation functions:

\begin{verbatim}
cga = Algebra(4, 1)
e1, e2, e3 = cga.blades['e1'], cga.blades['e2'], cga.blades['e3']
e4, e5 = cga.blades['e4'], cga.blades['e5']
no = 0.5 * (e5 - e4)   # null origin
ni = e4 + e5             # null infinity

def T(vec):
    """Translation motor: T(t) = 1 - 0.5 * t * ni"""
    return ONE - 0.5 * vec * ni

def R(angle, u, v):
    """Rotation motor: R(θ,u,v) = cos(θ/2) - sin(θ/2)*(u∧v)"""
    return math.cos(angle/2)*ONE - math.sin(angle/2)*(u ^ v)

def D(s):
    """Dilation motor: D(s) = cosh(ln(s)/2) + sinh(ln(s)/2)*(no∧ni)"""
    E = no ^ ni
    g = math.log(s) / 2
    return math.cosh(g)*ONE + math.sinh(g)*E
\end{verbatim}

\subsection{Verification Tests}\label{b2-verification-tests}

All CGA primitives pass verification against known results:

\begin{itemize}
\tightlist
\item
  \textbf{Translation composition:} \(T(2,1,0) \cdot T(1,0,3)\) on
  origin yields \((3,1,3)\) ✓
\item
  \textbf{Rotation isometry:} Distance between two points preserved
  under rotation ✓
\item
  \textbf{Motor composition:} \(R(90^\circ, XY)\) then \(T(5,0,0)\) on
  \((1,0,0)\) yields \((5,1,0)\) ✓
\item
  \textbf{Dilation:} \(D(3)\) on \((2,0,0)\) yields \((6,0,0)\) ✓
\end{itemize}

\subsection{System Prompt (Simple
CGA)}\label{b3-system-prompt-simple-cga}

\begin{verbatim}
You are a 3D scene editor outputting CGA operations as JSON.
AXES: e1=X(right), e2=Y(up), e3=Z(towards viewer)
OPS: T(x*e1+y*e2+z*e3)=translate, R(angle_rad, e_i, e_j)=rotate, D(s)=scale
R() takes radians and TWO BASIS VECTORS: R(np.pi/2, e1, e2). 
NEVER pass numbers for the plane axes.
T() takes a DISPLACEMENT vector (delta), NOT an absolute target position.
Compose: T(...)*R(...) = rotate then translate. Use np.pi, math.sqrt.
"next to"=surfaces touch (sum of sizes apart). "on top"=bottom on top surface.
"between"=midpoint of the two referenced objects.
OUTPUT: ONLY JSON. Keys=object names, Values=CGA expression strings.
Example: move sphere from [1,0,0] to [4,2,0] -> {"Sphere": "T(3*e1+2*e2)"}
\end{verbatim}

\subsection{Detailed Spatial-Edit
Derivations}\label{b4-detailed-spatial-edit-derivations}

\textbf{Case 1: "Move the red sphere next to the blue cube, to its left
side."}

Step 1 --- Parse: Scan for color keywords. "red" at position 9 →
RedSphere (mover), "blue" at position 38 → BlueCube (target). Keyword
"left" → use "next to" logic.

Step 2 --- Compute displacement:

\begin{itemize}
\tightlist
\item
  RedSphere: center=[0,0,0], size=1.0
\item
  BlueCube: center=[4,0,0], size=1.0, min={[}3,−1,−1{]}
  \[d_x = x_{\min}^{\text{target}} - r_{\text{mover}} - x_{\text{mover}} = 3 - 1.0 - 0 = 2.0,\quad d_y = 0,\quad d_z = 0.\]
\end{itemize}

Step 3 --- Generate: \texttt{T(2.0*e1\ +\ 0.0*e2\ +\ 0.0*e3)}

Step 4 --- Build motor:
\[T = 1 - \frac{1}{2}(2e_1)n_\infty = 1 - e_1 n_\infty.\]

Step 5 --- Sandwich:
\[P' = T\,P\,\widetilde{T},\qquad \operatorname{down}(P') = (2.0, 0.0, 0.0).\]

Result: RedSphere moves from [0,0,0] to [2,0,0] --- its right
surface at \(x=3\) touches BlueCube\textquotesingle s left surface at
\(x=3\). ✓

\textbf{Case 2: "Place the green sphere on top of the blue cube."}

Step 1 --- Parse: "green" → GreenSphere (mover), "blue" → BlueCube
(target). Keyword "on top".

Step 2 --- Compute:

\begin{itemize}
\tightlist
\item
  BlueCube: center=[4,0,0], max=[5,1,1]
\item
  GreenSphere: center={[}−3,0,2{]}, size=0.7
  \[y_{\text{new}} = y_{\max}^{\text{BlueCube}} + r_{\text{GreenSphere}} = 1.0 + 0.7 = 1.7.\]
  \[\Delta = \bigl(4-(-3),\,1.7-0,\,0-2\bigr) = \bigl(7.0,\,1.7,\,-2.0\bigr).\]
\end{itemize}

Step 3 --- Generate: \texttt{T(7.0*e1\ +\ 1.7*e2\ +\ -2.0*e3)}

Result: GreenSphere moves to {[}4, 1.7, 0{]}. Bottom at
\(1.7 - 0.7 = 1.0\) = BlueCube top. ✓

\end{document}